\documentclass[12pt]{iopart}

\expandafter\let\csname equation*\endcsname\relax
\expandafter\let\csname endequation*\endcsname\relax
\usepackage{iopams}
\usepackage{graphicx}
\usepackage{amssymb}
\usepackage{amsmath}
\usepackage{graphicx}
\usepackage[usenames]{color}
\usepackage{url}
\usepackage{cite}

\begin{document}

\title[Finite size scaling in crossover among different random matrix ensembles]{Finite size scaling in crossover among different random matrix ensembles in microscopic lattice models}

\author{ Ranjan Modak$^1$ and Subroto Mukerjee$^{1,2}$}
\address{$^1$ Department of Physics, Indian Institute of Science, Bangalore 560 012, India}
\address{$^{2}$ Centre for Quantum Information and Quantum Computing, Indian Institute of Science,
Bangalore 560 012, India}

\begin{abstract}
Using numerical diagonalization we study the crossover among different random matrix ensembles [Poissonian, Gaussian Orthogonal Ensemble (GOE), Gaussian Unitary Ensemble (GUE) and Gaussian Symplectic Ensemble (GSE)] realized in two different microscopic models. The specific diagnostic tool used to study the crossovers is the level spacing distribution. The first model is a one dimensional lattice model of interacting hard core bosons (or equivalently spin 1/2 objects) and the other a higher dimensional model of non-interacting particles with disorder and spin orbit coupling. We find that the perturbation causing the crossover among the different ensembles scales to zero with system size as a power law with an exponent that depends on the ensembles between which the crossover takes place. This exponent is independent of microscopic details of the perturbation. We also find that the crossover from the Poissonian ensemble to the other three is dominated by the Poissonian to GOE crossover which introduces level
repulsion while the crossover from GOE to GUE or GOE to GSE associated with symmetry breaking introduces a subdominant contribution. We also conjecture that the exponent is dependent on whether the system contains interactions among the elementary degrees of freedom or not and is independent of the dimensionality of the system.

\end{abstract}

\pacs{02.30.Ik, 05.30.-d,05.45.Mt}


\maketitle
\section{Introduction}
Random Matrix Theory (RMT)~\cite{mehta} was first applied to physical systems in the context of nuclear physics~\cite{dyson1962statistical,mehta1960statistical} and has found application in condensed matter physics especially in the study of disordered systems. It has also been shown to play a crucial role in understanding how isolated quantum systems thermalize~\cite{montambaux.1993}. In contrast to classical mechanics
there is no notion of a phase space in quantum mechanics. Hence, the concept of quantum ergodicity
(thermalization) is not very well understood, and is presently a very active area of research
\cite{deutsch1991,srednicki1994,rigol.2008}.It is believed that isolated quantum systems that thermalize generally do not have dynamics strongly constrained by conservation laws and are thus not integrable. These
non-integrable systems can be characterized by random Matrix ensembles depending on the symmetry of their Hamiltonians.

Integrable models have infinite conserved quantities in the thermodynamic limit~\cite{shastry.1986}, as a consequence of which they display no level repulsion and obey a Poissonian level spacing distribution given by $P(s)=\exp(-s)$,
where $s$ is energy spacing measured in units of the mean level spacing. In contrast a non-integrable system has a finite
number of conserved quantities even in the thermodynamic limit. Once, one has accounted for the corresponding symmetries, the rest of the energy spectrum displays level repulsion with $P(s)\to 0$ as $s\to 0$.
Depending on the symmetries of the system, $P(s)$ can have the following forms:

\begin{enumerate}
\item $P(s)=\pi s/2\exp(-\pi s^{2}/4)$ for the Gaussian Orthogonal Ensemble (GOE),
\item $P(s)=32s^{2}/\pi^{2}\exp(-4s^{2}/\pi)$ for the Gaussian Unitary Ensemble (GUE) (where time reversal symmetry is broken)
\item $P(s)=(2^{18}/3^{6}\pi^{3})s^{4}\exp(-(64/9\pi)s^{2})$ for Gaussian Symplectic
Ensemble(GSE) (where time reversal symmetry is preserved but spin rotation symmetry is
broken).
\end{enumerate}

In the presence of disorder, the situation is different. The disorder renders the system non-integrable by destroying conservation laws that may have existed in its absence. However, it is possible that for a sufficiently strong value of disorder, the level spacing distribution is Poissonian indicative of localization in the system. For non-interacting disordered systems (without spin orbit coupling), it is known that in one
and two dimensions, even an infinitesimal amount of disorder is sufficient to localize all
states in the thermodynamic limit~\cite{anderson.1958,ramakrishnan.1979}. In three dimension even in the thermodynamic limit one needs to have a finite amount of disorder to localize all states. Significantly less is understood about the nature of localization for
interacting disordered systems. The one parameter scaling theory for non-interacting systems ~\cite{ramakrishnan.1979} is expected not to apply in this case and there is much debate over whether there is a finite amount of disorder required for localization and what its dependence on interaction strength and dimensionality is~\cite{huse.2007,song.2000}. We do not attempt to join the debate in this paper focussing instead on systems with weak enough disorder that localization does not occur. In a previous work with Ramaswamy we investigated how integrability in a one-dimensional interacting system is destroyed by perturbations~\cite{modak.2013}. We found that the scale of the perturbation that caused a crossover to non-integrability goes to zero with increasing system size as a power law in the system size whose exponent is independent of microscopic details. We conjectured that the value of this exponent was dependent only on the random matrix ensemble describing the non-integrable system which was of the
GOE type. In this paper, we elaborate on that claim by studying systems described by different random matrix ensembles and and investigate the crossovers among them. We also investigate the effect of dimensionality on the finite-size scaling of the perturbation that causes the crossover. To this end, it is most convenient for us to look at models which have disorder in addition to interactions.

In this paper, we have looked at two disordered models, 1) a one dimensional interacting model of
hard-core bosons and 2) a three dimensional model of non-interacting particles with Spin-Orbit-Coupling(SOC). These models allow us to realize phases with Poissonian, GOE, GUE and GSE level spacing
distributions and thus study the crossovers among them. Our main result is that the scale of perturbations responsible for the crossovers among the different classes of systems goes to zero with increasing system size as a power law, with an exponent that appears to depend only on the random matrix ensembles of the classes independent of microscopic details.

\section{Models}

\subsection{Model I: One dimensional interacting system with disorder}

We consider the one-dimensional Heisenberg spin-1/2 chain containing $N$ spins with
random on site magnetic field in the $z$-direction.
The Hamiltonian for this system is:

\begin{equation}
H=\sum_{j=1}^{N}J\bold{S_{j}.S_{j+1}}+h_{j}S_{j}^{z},
\label{Eqn:Hamiltonian}
\end{equation}
where $S_j$ is the spin operator at site $j$ and $J$ is the nearest neighbor exchange constant.
$h_{j}$ is a random magnetic
field in the $z$ direction, which is uniformly distributed in the interval $[-h/2,h/2]$.
\par
For this model $T_{0}S_{j}T_{0}^{-1}=-S_{j}$, where $T_0$ is the time reversal operator and therefore,
\begin{equation}
T_{0}HT_{0}^{-1}=\sum_{j=1}^{N}J\bold{S_{j}.S_{j+1}}-h_{j}S_{j}^{z} \neq H.
\label{Eqn:timereversal Hamiltonian}
\end{equation}
The  presence of a magnetic field breaks the time reversal
symmetry.  However, the antiunitary operator $T=e^{i\pi S^{x}}T_{0}$ commutes with $H$.
The operator $e^{i\pi S^{x}}$ reverses the
sign of $S_{j}^{y}$ and $S_{j}^{z}$ but not of $S_{j}^{x}$~\cite{Haake}.
\par
 $H$ thus preserves an unconventional time reversal symmetry and the level spacing distribution for $H$ is GOE type in the presence of a large enough random magnetic field that is not so large as to localize all states.
\par
Introducing a three site interaction~\cite{avishai.2002}, the Hamiltonian becomes,

\begin{equation}
H=\sum_{j=1}^{N}J\bold{S_{j}.S_{j+1}}+h_{j}S_{n}^{j} +J_{T}\bold{S_{j}.[S_{j+1}\times S_{j+2}]},
\label{Eqn:Hamiltonian_broken T}
\end{equation}
which breaks time reversal symmetry and unlike for the Hamiltonian of Eqn.~\ref{Eqn:Hamiltonian}, this time reversal symmetry violation can not be compensated by any anti-unitary spin reversal operator. The three spin term can be written as,
\begin{eqnarray}
J_{T}\bold{S_{j}.[S_{j+1}\times S_{j+2}]}&=&i J_{T}S_{j}^{z}\epsilon_{jkl}S_{k}^{+}S_{l}^{-} \nonumber \\
&=&i J_{T}S_{j}^{z}S_{j+1}^{+}S_{j+2}^{-}-i J_{T}S_{j}^{z}S_{j+2}^{+}S_{j+1}^{-}
\label{Eqn:triple product term}
\end{eqnarray}

 and using a Holstein and Primakoff transformation~\cite{cazalilla2011} this model can be mapped onto one of hardcore bosons. The spin operators in terms of the bosonic operators are
\begin{eqnarray}
S_{j}^{+}&=&b_{j}^{\dag}\sqrt{1-n_{j}}\nonumber\\
S_{j}^{-}&=&\sqrt{1-n_{j}}b_{j}\nonumber\\
S_{j}^{z}&=&n_{j}-1/2
\end{eqnarray}
where, $b_{j}^{\dag}$ and $b$ are the creation and annihilation operators for the hard core bosons and
$n_{j}=b_{j}^{\dag}b_{j}$ is the number
density operator.
The resultant Hamiltonian, which we have studied is thus,

\begin{eqnarray}
H=t\sum_{j}{b_{j}^{\dag}b_{j+1}+h.c.}&+&V\sum_{j}{n_{j}n_{j+1}}+\sum_{j}{h_{j}n_{j}} \nonumber \\
  &+&t'\sum_{j}{i n_{j}b_{j+1}^{\dag}b_{j+2}-in_{j}b_{j+2}^{\dag}b_{j+1}}
\label{Eqn:hamiltonian_hardcore}
\end{eqnarray}

If $t'=h_{j}=0$, the model reduces to the integrable $t-V$ model of Hardcore boson~\cite{marcos.2009}. Non-zero
$h_{j}$ makes the model non-integrable and also destroys its lattice translational symmetry.
We have used exact diagonalization techniques to obtain all the eigenvalues of the Hamiltonian
and have able to reach up to $N=16$ sites at half filling. The results we report have been averaged over different realizations of disorder to achieve convergence.

\subsection{Model II: Three dimensional disordered model with spin orbit coupling (SOC)}

We consider a three-dimensional disordered system ~\cite{evangelou.1995,reza.2010} with SOC
on a cubic lattice. The Hamiltonian of this model is given by:
\begin{equation}
 H=\sum_{i\sigma}h _{i}c^\dagger_{i\sigma} c_{i\sigma} + (\sum_{<ij>}\sum _{\sigma \sigma '}
 V_{i,j;\sigma,\sigma'}c^\dagger_{i\sigma} c_{j\sigma}+ h.c.),
\label{Eq:socham}
\end{equation}
where $i$ labels the sites of the lattice and $\sigma$ labels the spin. $<ij>$ labels nearest neighbor pairs $i$ and $j$. $h_{i}$ is $a$ random on-site potential which does not contain the spin index and thus does not violate time reversal invariance. $h_{i}$ is chosen from the interval $(-h/2,h/2)$ of
uniformly distributed random variables.
The nearest neighbor hopping $V_{i,j}$ has the following form:
\begin{eqnarray}
V_{i,j}=
\quad
\begin{pmatrix}
1+i\mu V^{z} & \mu V^{y}+i\mu V^{x} \\
-\mu V^{y}+i\mu V^{x} & 1-i\mu V^{z}
\end{pmatrix}
\quad
\label{Eq:intform}
\end{eqnarray}
where, $\mu$ is the spin orbit coupling strength and $V^{x}$,$V^{y}$ and $V^{z}$ are independent uniform random variables
taken from the interval $(-1/2,1/2)$. For nonzero $\mu$, this model breaks spin rotation symmetry and hence its eigenvalues give a GSE type level spacing distribution. When $\mu$ is zero, the model belongs to the
GOE class. A large value of $h$ will cause localization and the again energy eigenvalues will
obey a Poissonian level spacing distribution. We will restrict ourself to a region where model exhibits no localization.

A study of the crossover from Poissonian to GSE level spacing statistics can be studied even in two dimensions for the model of Eqn.~\ref{Eq:socham}. However, the crossover from Poissonian to GOE statistics cannot be examined in this model in two dimensions since this requires the SOC to be turned off which will cause localization (and no phase with GOE statistics) for any amount of disorder.

Like the System described by Eqn.~\ref{Eqn:Hamiltonian}, this model too has disorder and hence no lattice translation symmetry. We thus write down the Hamiltonian in a real space basis to perform exact diagonalization to obtain all the energy eigenstates. We have been able to diagonalize systems with $L^3$ sites and a maximum value of $L=15$. Once again, we average over different realizations of disorder to obtain good statistics for the quantities of interest.

\section{Methodology}

\subsection{Poissonian to GOE}

\begin{figure}
\centering
\begin{tabular}{cc}
\includegraphics[width=3.0in]{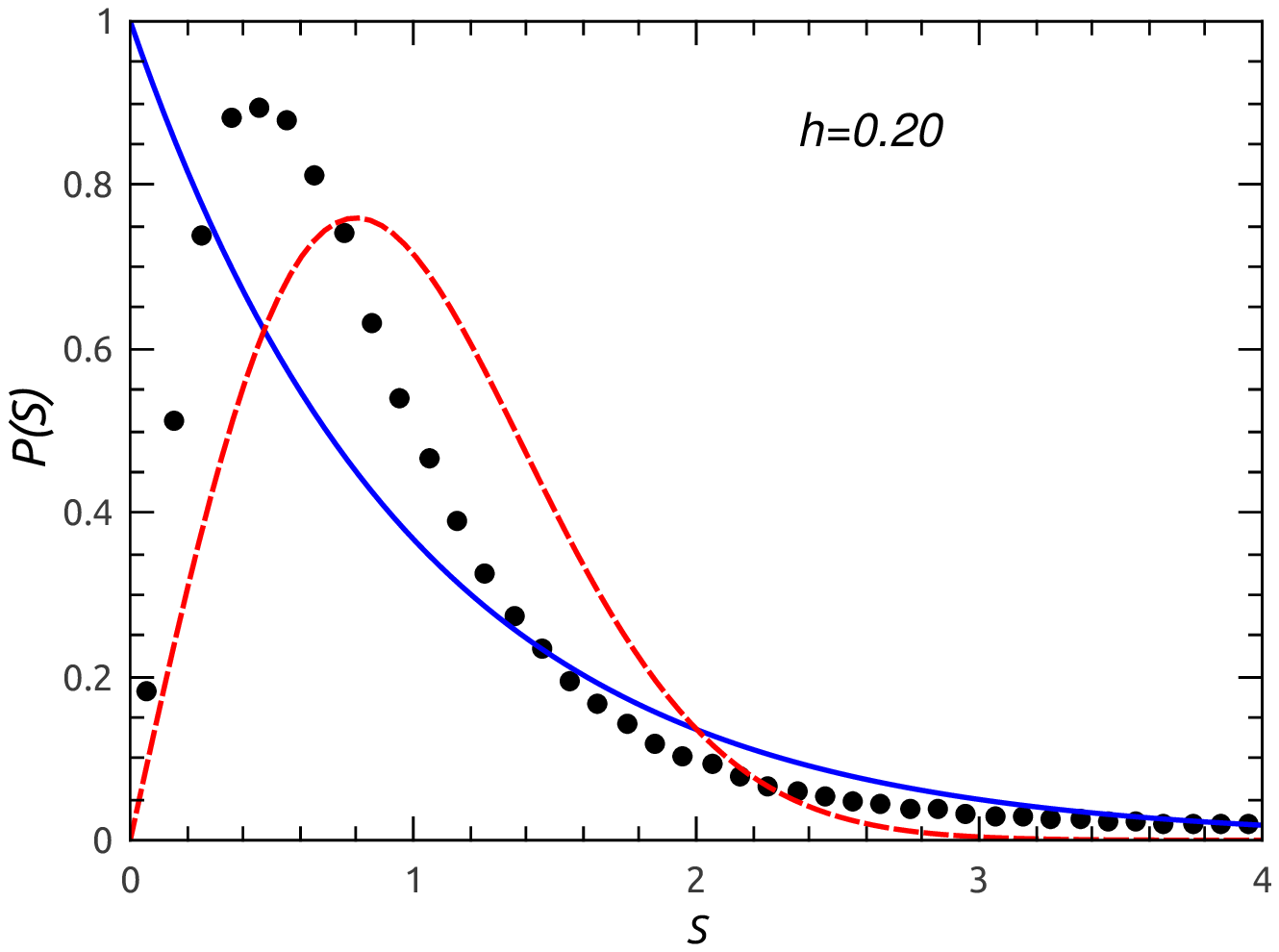} &
\includegraphics[width=3.0in]{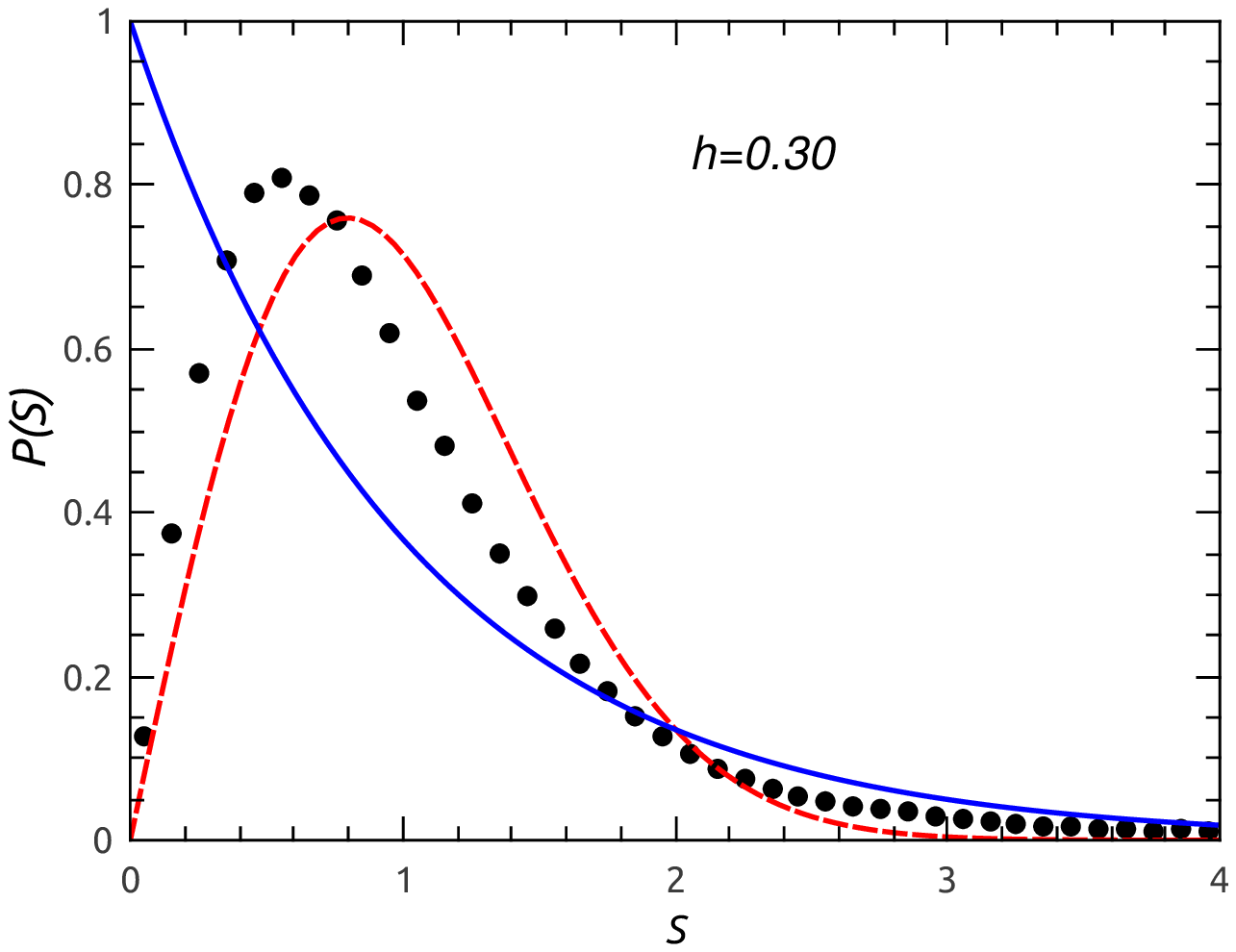} \\
\includegraphics[width=3.0in]{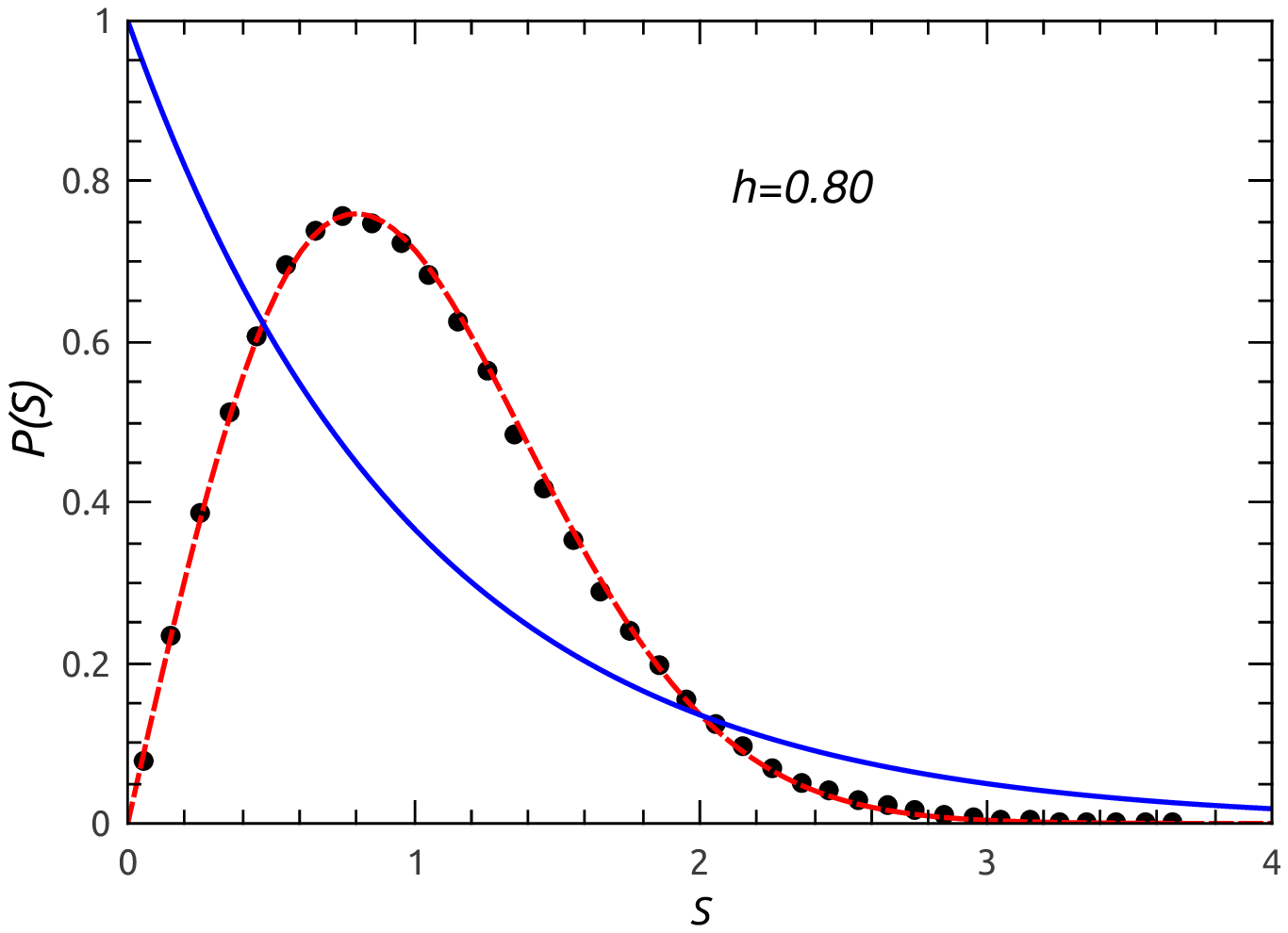} &
\includegraphics[width=3.0in]{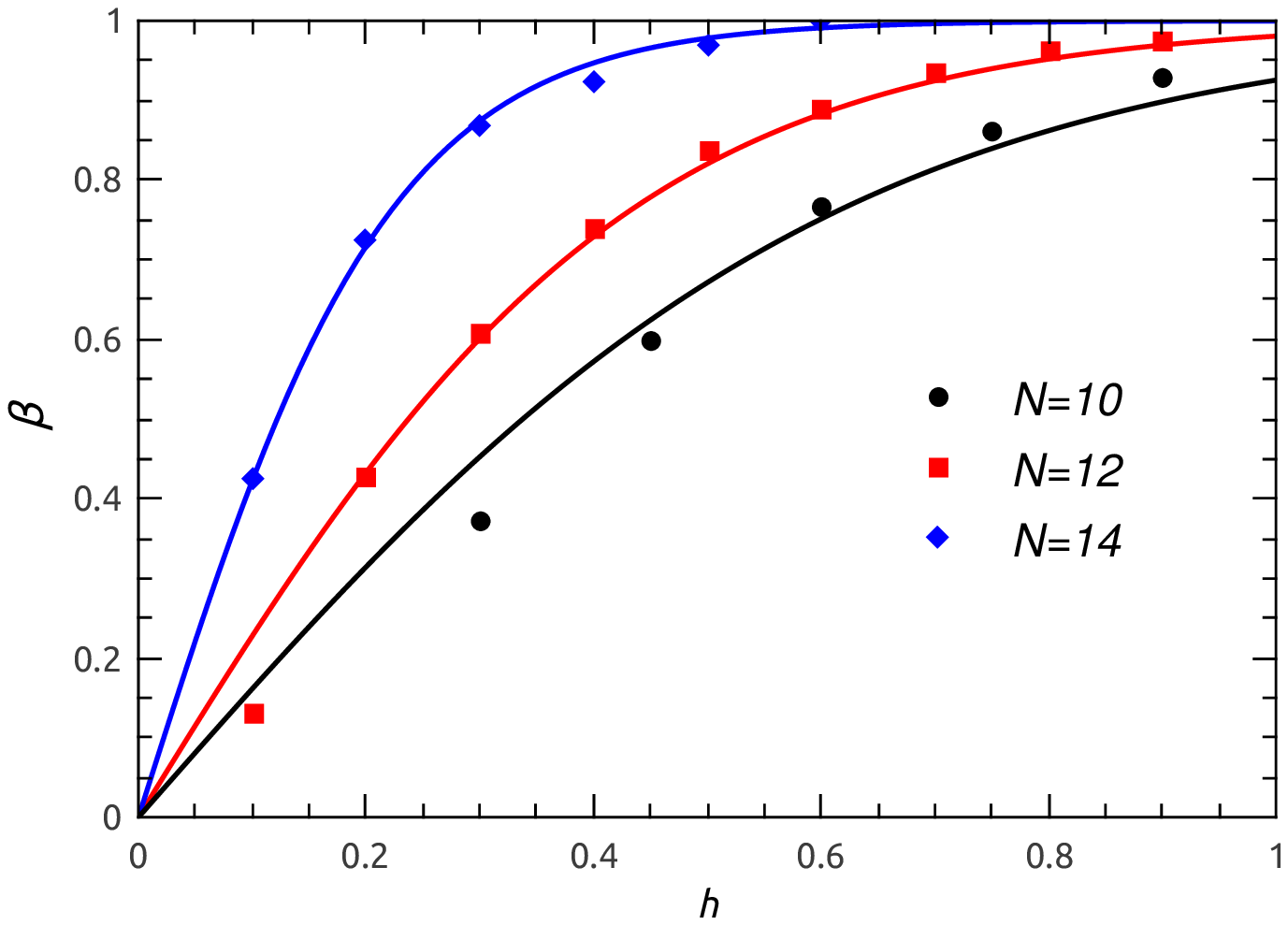}
\
\end{tabular}
\caption{For Model I(A) change in level spacing distribution from Poissonian to GOE for $N=12$
(at half filling) for $t=-1$
and $V=2$ and $t'=0$ as $h$ is increased. The dashed line is a fit to the GOE distribution and the solid line is to a Poissonian distribution.(B) The variation of $\beta$  defined in Eq.~\ref{Eq:brody} with $h$
 for $N=10,12,14$. The solid lines are fits to the function $\tanh(h/h_{cr})$,
 where $h_{cr}$ is a measure of
 the crossover value.}
\label{Fig:level spacing Poi-GOE}
\end{figure}

Setting $h=t'=0$ in Eqn.~\ref{Eqn:hamiltonian_hardcore} we obtain an integrable model of hard-core bosons with Poissonian level spacing statistics~\cite{modak.2013}. Increasing the value of $h$ keeping $t'=0$, a crossover from Poissonian to GOE level spacing statistics is observed as shown in Figure~\ref{Fig:level spacing Poi-GOE}.

\begin{figure}
\centering
\begin{tabular}{cc}
\includegraphics[width=3.0in]{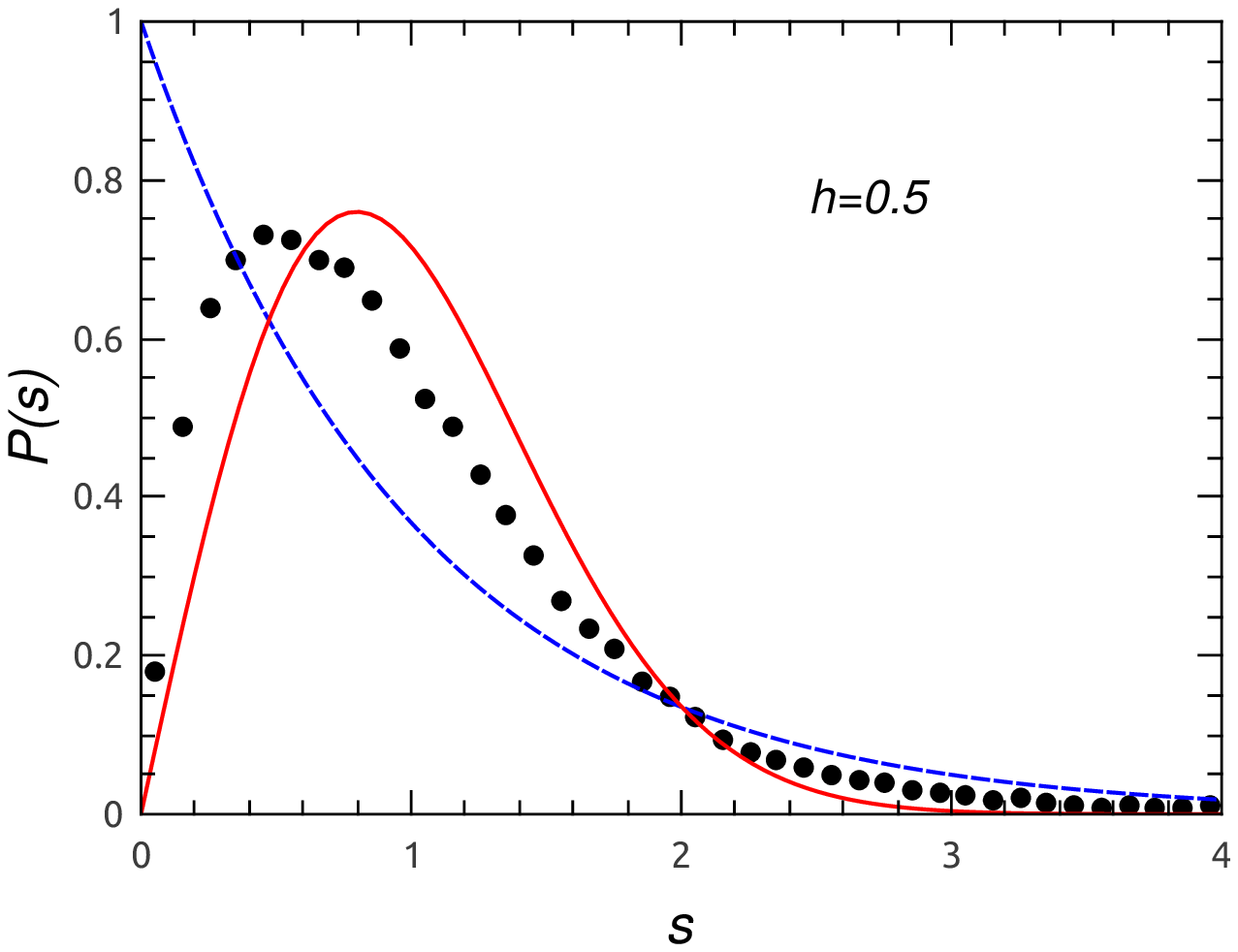} &
\includegraphics[width=3.0in]{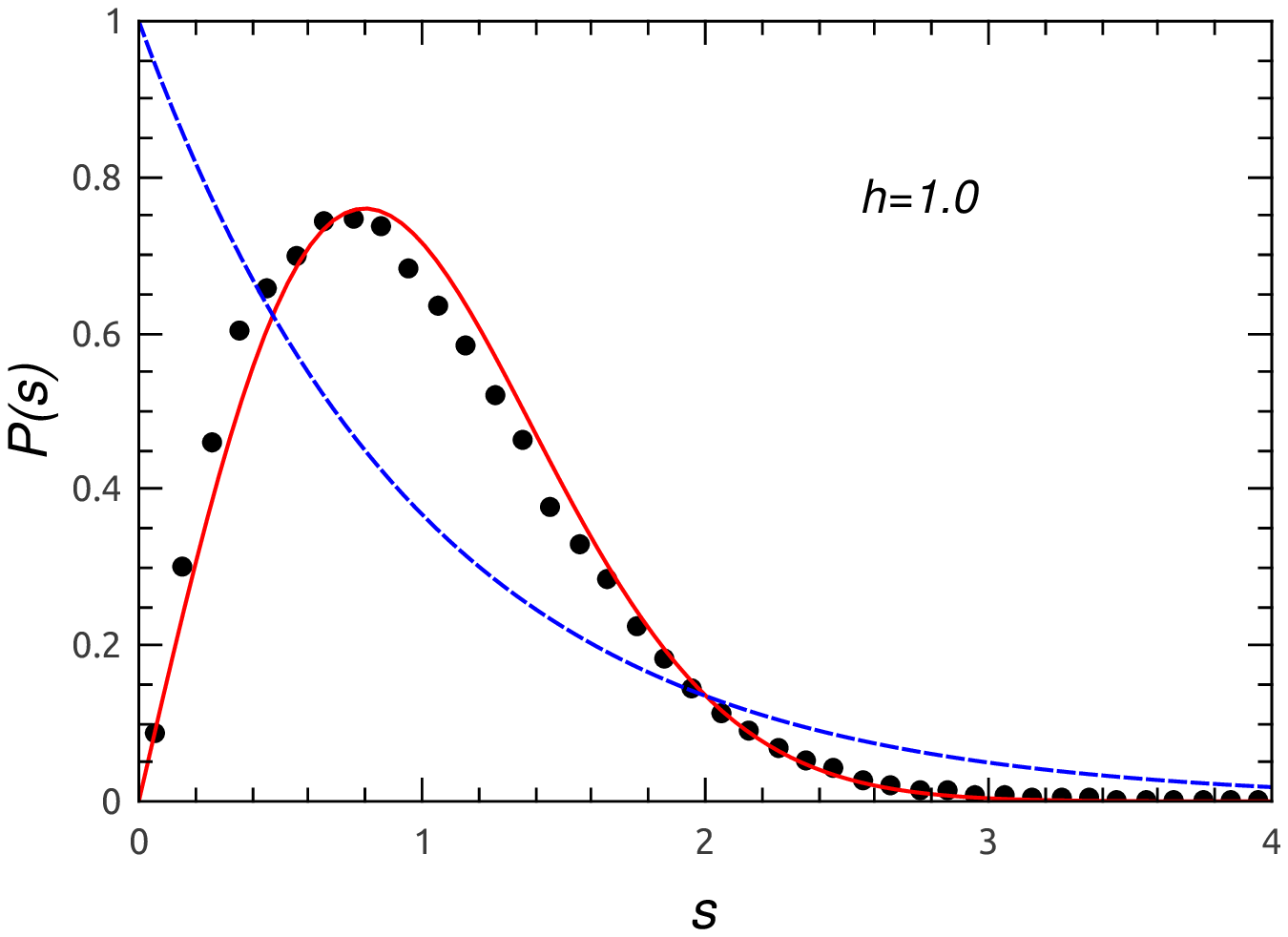} \\
\includegraphics[width=3.0in]{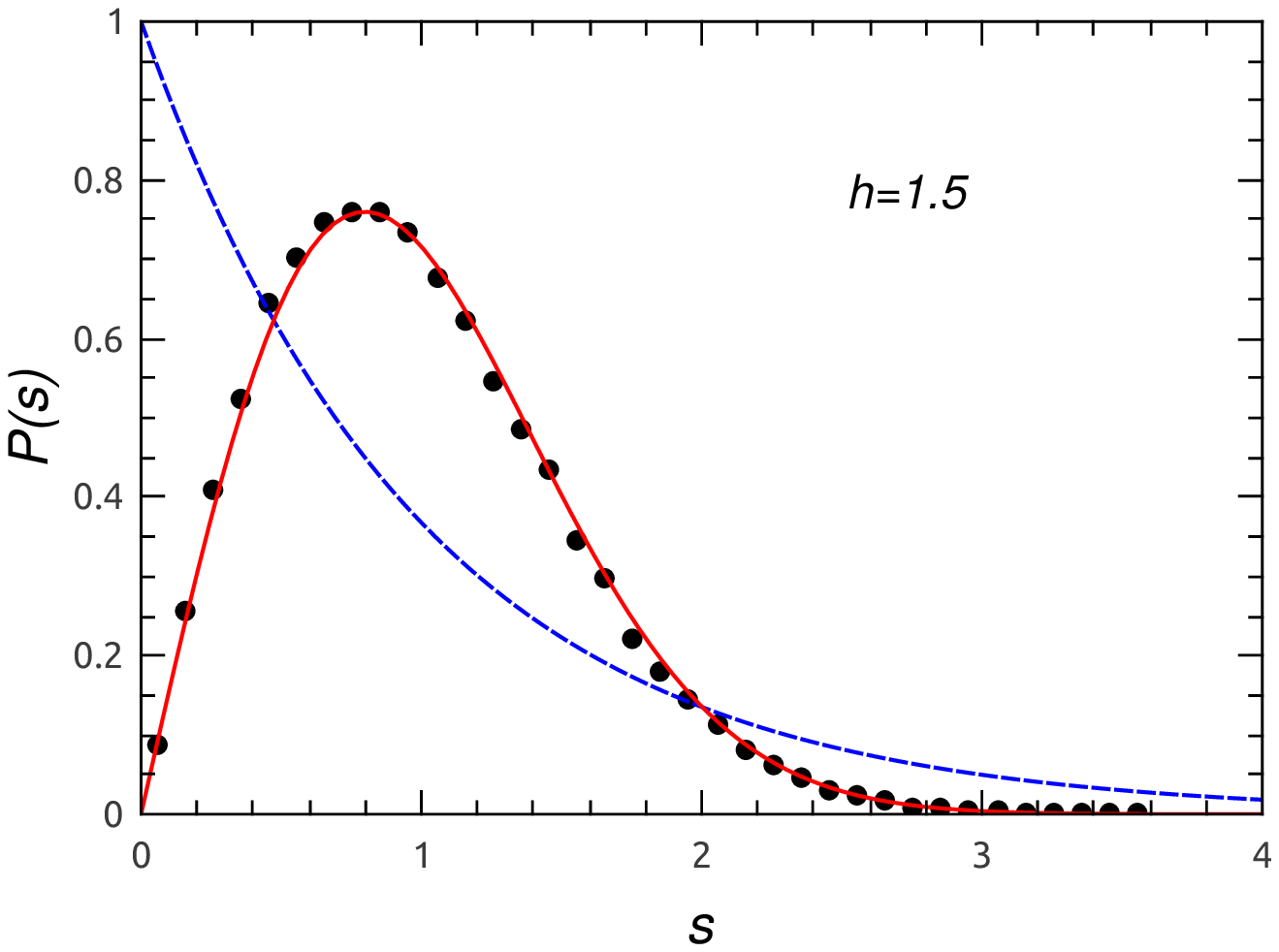} &
\includegraphics[width=3.0in]{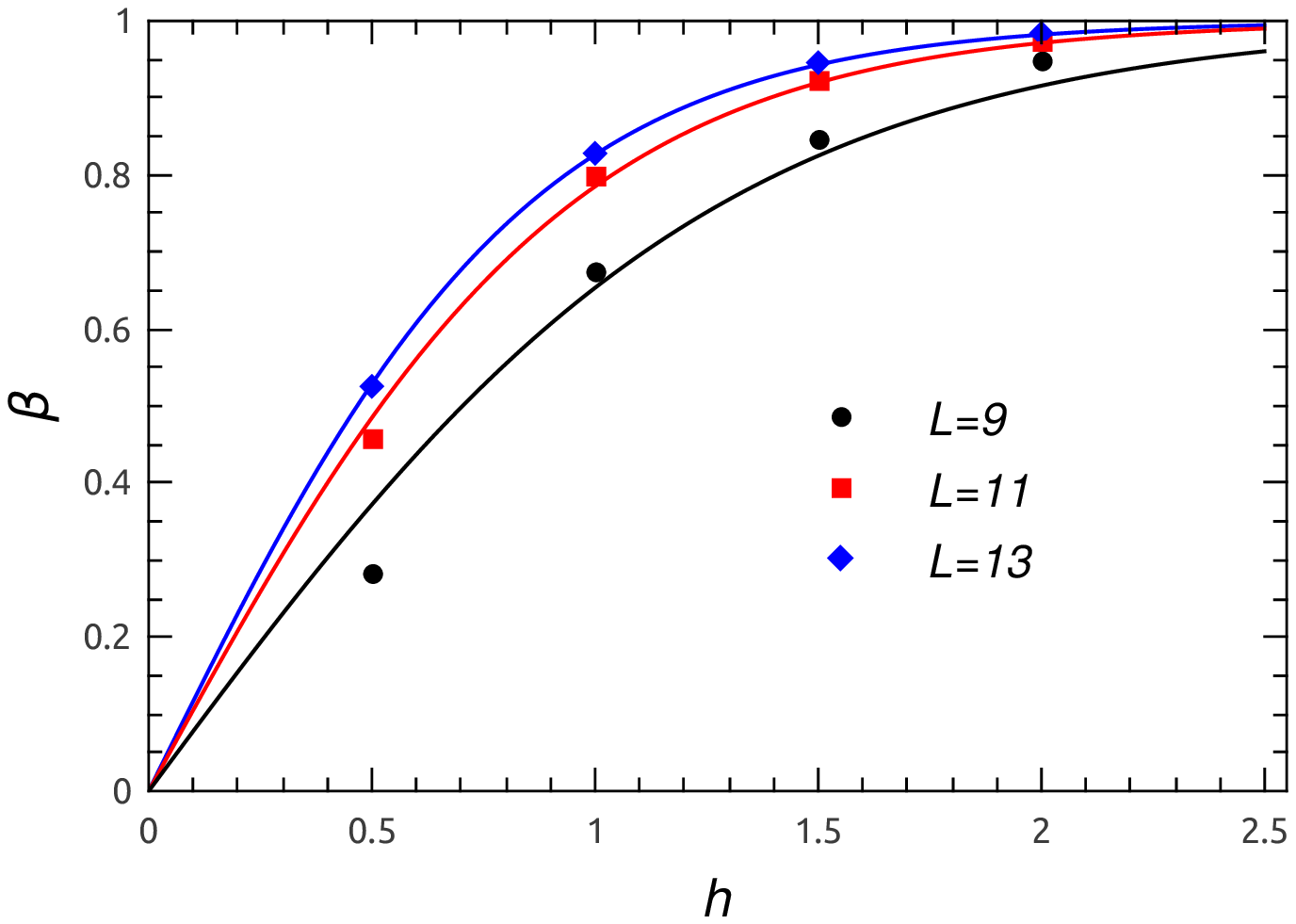}
\end{tabular}
\caption{(A)Level spacing distribution $P(s)$ for the 3D non-interacting disordered
model ($\mu=0$) for $L=13$. The values of the integrability
breaking parameter $h$ are 0.5,1.0,1.5. The dashed line is a fit to the Poissonian
distribution and the solid line to the GOE distribution.
(B) The variation of $\beta$ with $h$ and with a fit to the function $\tanh(h/h_{cr})$ for $L=9,11,13$}
\label{Fig:poi_goe_3D}
\end{figure}

For Model II, one can see a crossover from Poisson to GOE level spacing statistics as we increase the value of $h$ in Eqn.~\ref{Eq:socham} provided $\mu$ (the SOC strength) in Eqn.~\ref{Eq:intform} is
set to zero. This is shown in Figure~\ref{Fig:poi_goe_3D}.
The intermediate distribution can be to a Brody distribution~\cite{brody.1981}.

\begin{equation}
P(s)=(\beta+1)bs^{\beta}\exp(-bs^{\beta+1})
\label{Eq:brody}
\end{equation}
 where $b=(\Gamma[\frac{\beta+2}{\beta+1}])^{\beta+1}$. Eqn.~\ref{Eq:brody} interpolates between the Poissonian ($\beta=0$)
and GOE ($\beta=1$) distributions. We then plot $\beta$ as a function of the integrability breaking parameter (which is the disorder strength $h$ for both models) and make a fit to the function $\tanh(h/h_{cr})$. The parameter, $h_{cr}$ is the crossover value of disorder strength for a given system size~\cite{rabson.2004,modak.2013}.

\subsection{Poissonian to GUE}
\begin{figure}
\centering
\begin{tabular}{cc}
\includegraphics[width=3.0in]{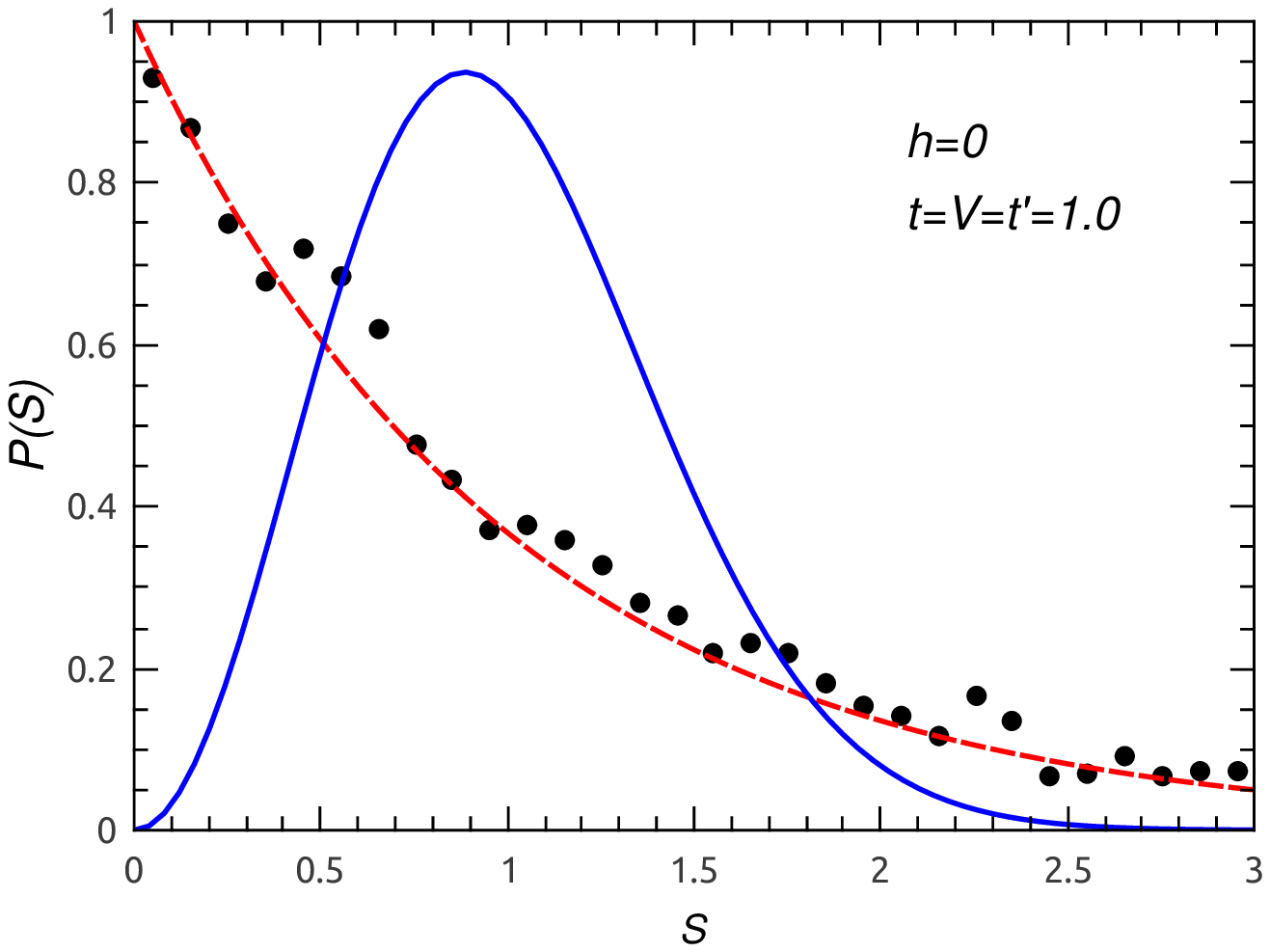} &
\includegraphics[width=3.0in]{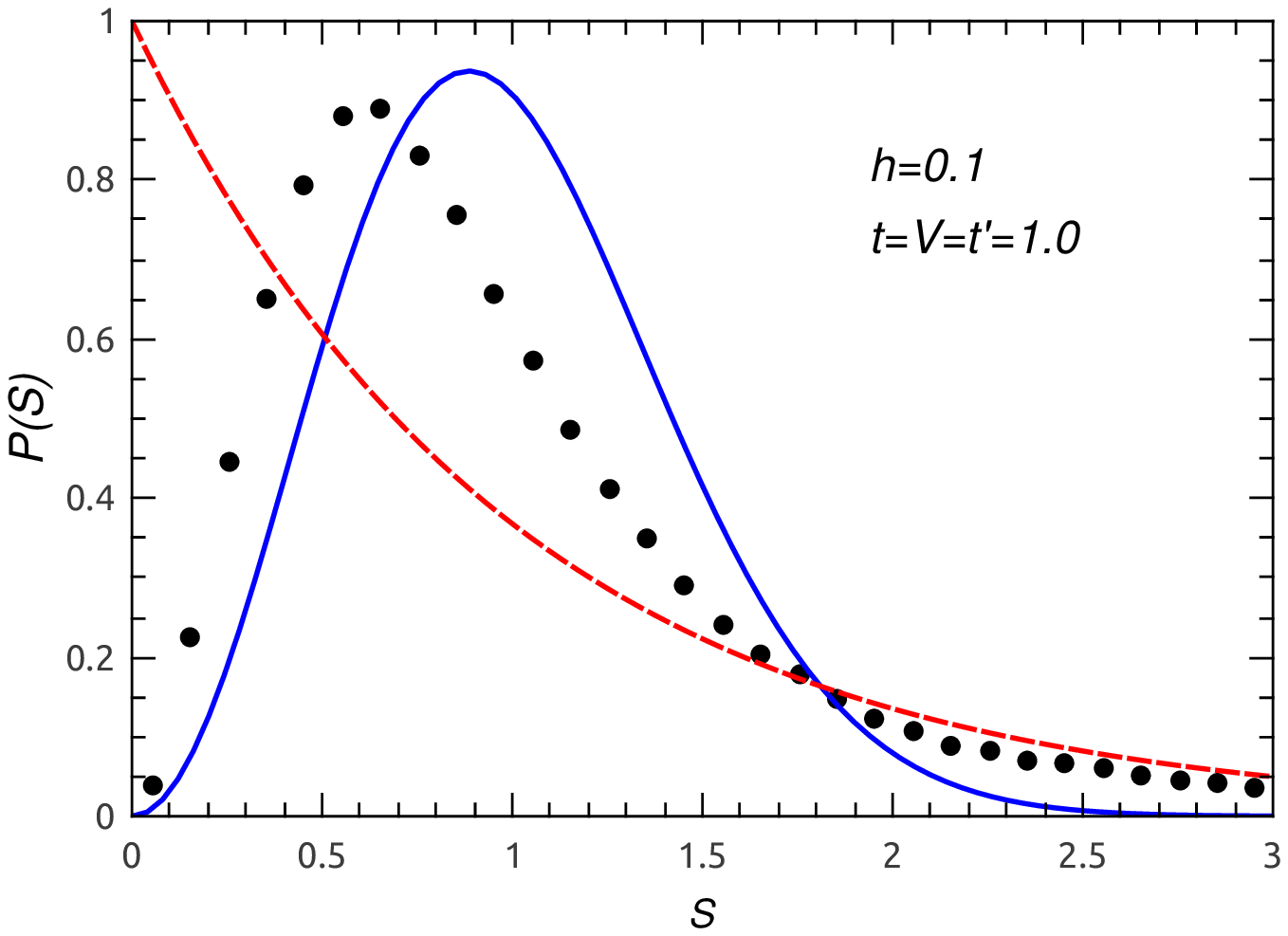} \\
\includegraphics[width=3.0in]{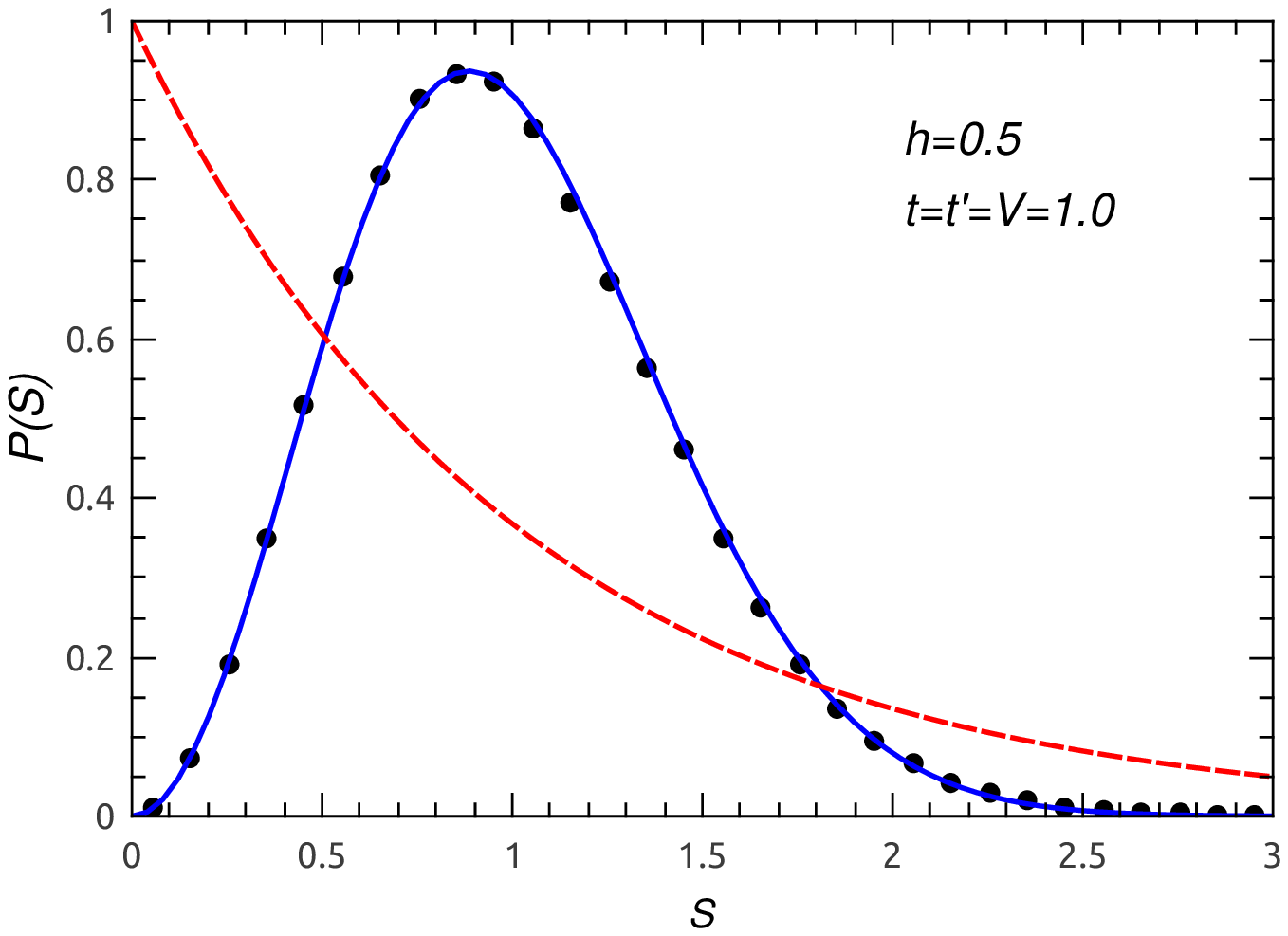} &
\includegraphics[width=3.0in]{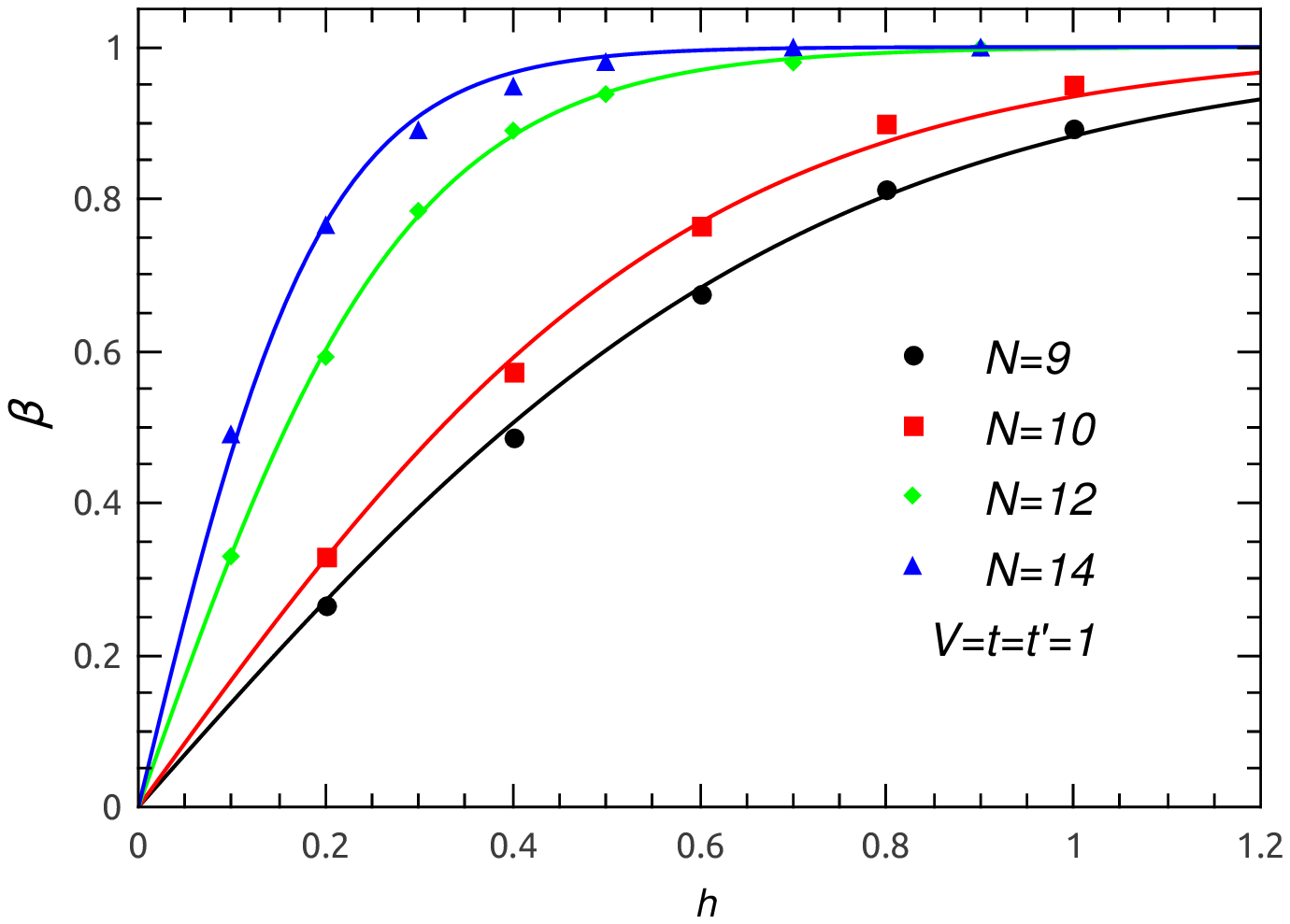}
\end{tabular}
\caption{For Model I (A), the change in level spacing distribution from Poissonian to GUE for $N=14$ (at half filling) for $t=V=t'=1$ as
$h$ is increased. The dashed line is a fit to the Poissonian distribution and the solid line is to GUE.
(B) $\beta$ as a function of $h$ for N=9,10,12,14 and fit to the function $\tanh(h/h_{cr})$.}
\label{Fig:level spacing poi-GUE}
\end{figure}

We can obtain a Poisson-GUE cross over only for Model I since for Model II, the on site random potential is spin independent and does not break time reversal symmetry.
In case of model I, if we switch on $h$ and $t'$ both simultaneously or while keeping $t$,$V$,$t'$
at some finite value increase $h$ slowly, we csee that $P(s)$ changes from Poissonian to GUE as
shown in Figure~\ref{Fig:level spacing poi-GUE}.
For $t' \neq 0$ and $h=0$, the model is not integrable in usual sense it still displays Poissonian level spacing statistics as can be seen in Figure~\ref{Fig:level spacing poi-GUE}(it is quasi-integrable). Values of $P(s)$ for level spacing distributions intermediate to Poissonian and GUE can be fit to a Brody form like in the previous case but this time given by

\begin{eqnarray}
 P(s)=(1+\beta)a^{2}s^{2\beta}exp(-as^{\beta+1})
\label{Eq:brody_GUE}
\end{eqnarray}
where, $a=(\Gamma[\frac{1+2\beta}{1+\beta}])^{-1-\beta}$. $\beta=0(1)$ for Poissonian(GUE) level spacing distributions. Again, like in the previous case we have plotted $\beta$ as a function
of $h$ and fit to the function $\tanh(h/h_{cr})$ to obtain the crossover value of disorder strength for different system sizes.

\subsection{GOE to GUE}
As mentioned earlier time reversal symmetry is preserved in model II, hence the GOE to GUE crossover cannot
be observed in it. However for model I, as shown in Figure~\ref{Fig:level spacing GOE-GUE}
keeping $h$ fixed at some finite value (taken to be $h=4$) and $t'=0$ such that the GOE distribution is observed,
 we can easily obtain the GOE to GUE crossover by increasing $t'$. In this crossover region, $P(s)$ obeys the following Brody like one parameter functional form:

\begin{figure}
\centering
\begin{tabular}{cc}
\includegraphics[width=3.0in]{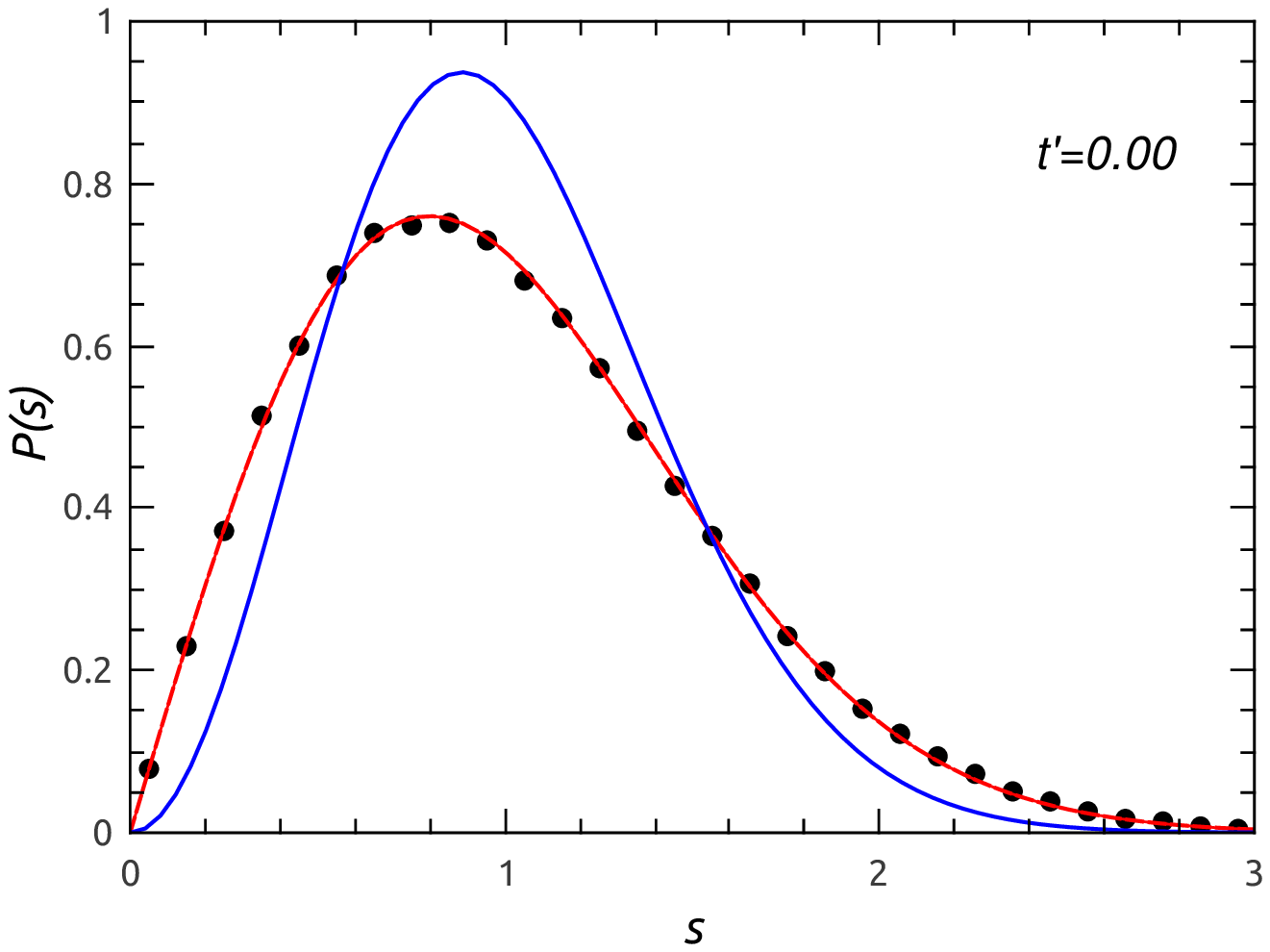} &
\includegraphics[width=3.0in]{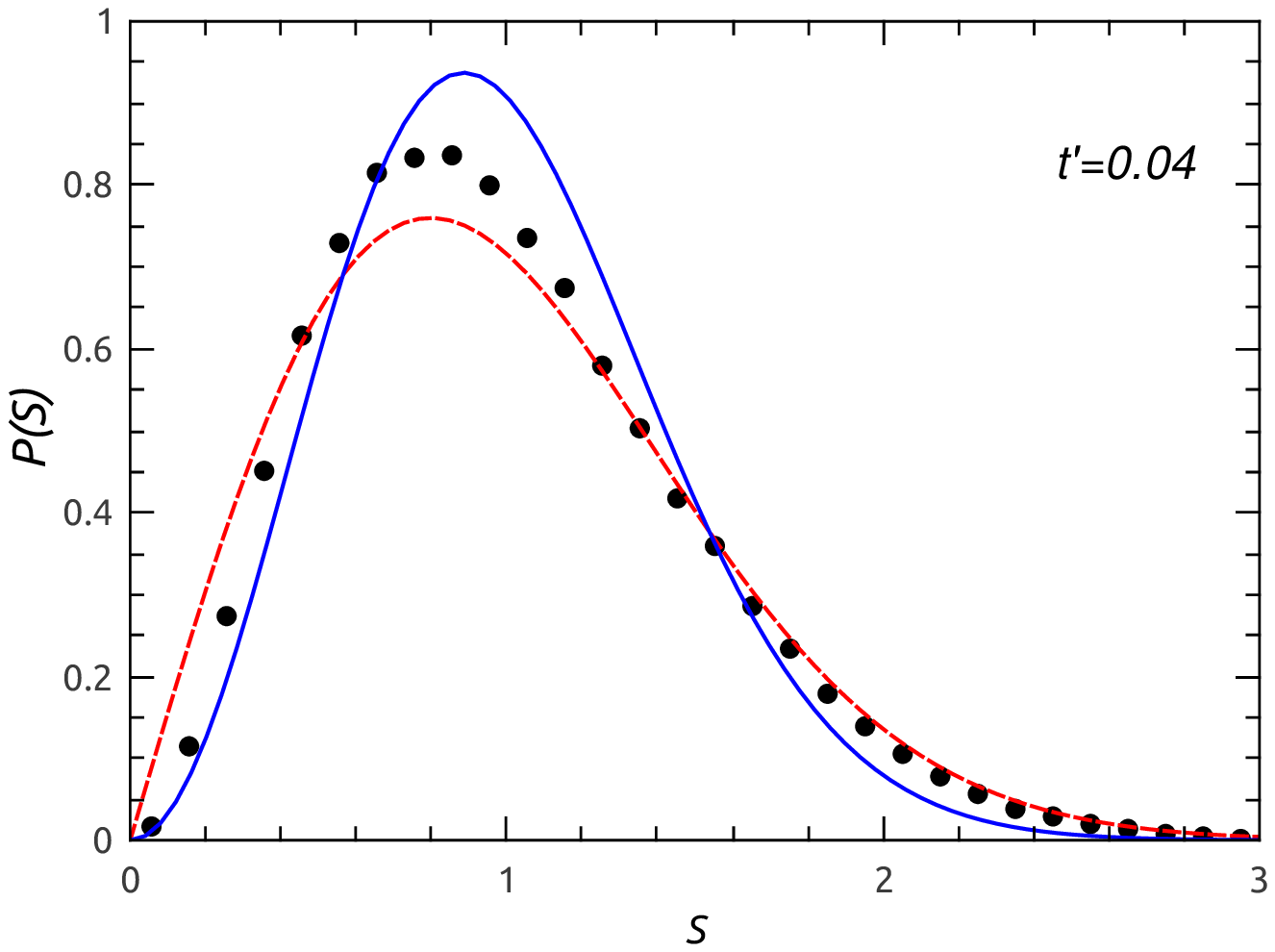} \\
\includegraphics[width=3.0in]{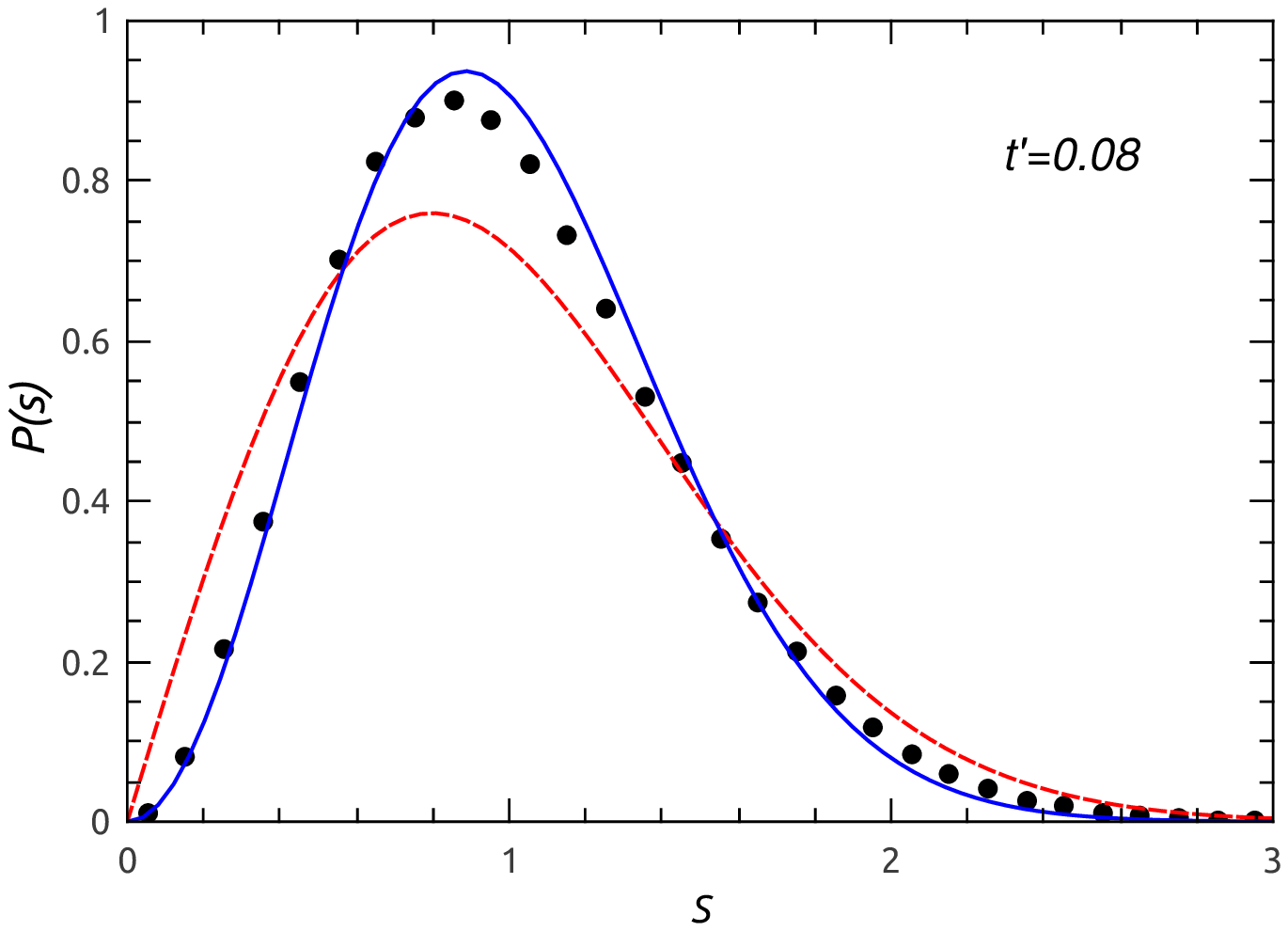} &
\includegraphics[width=3.0in]{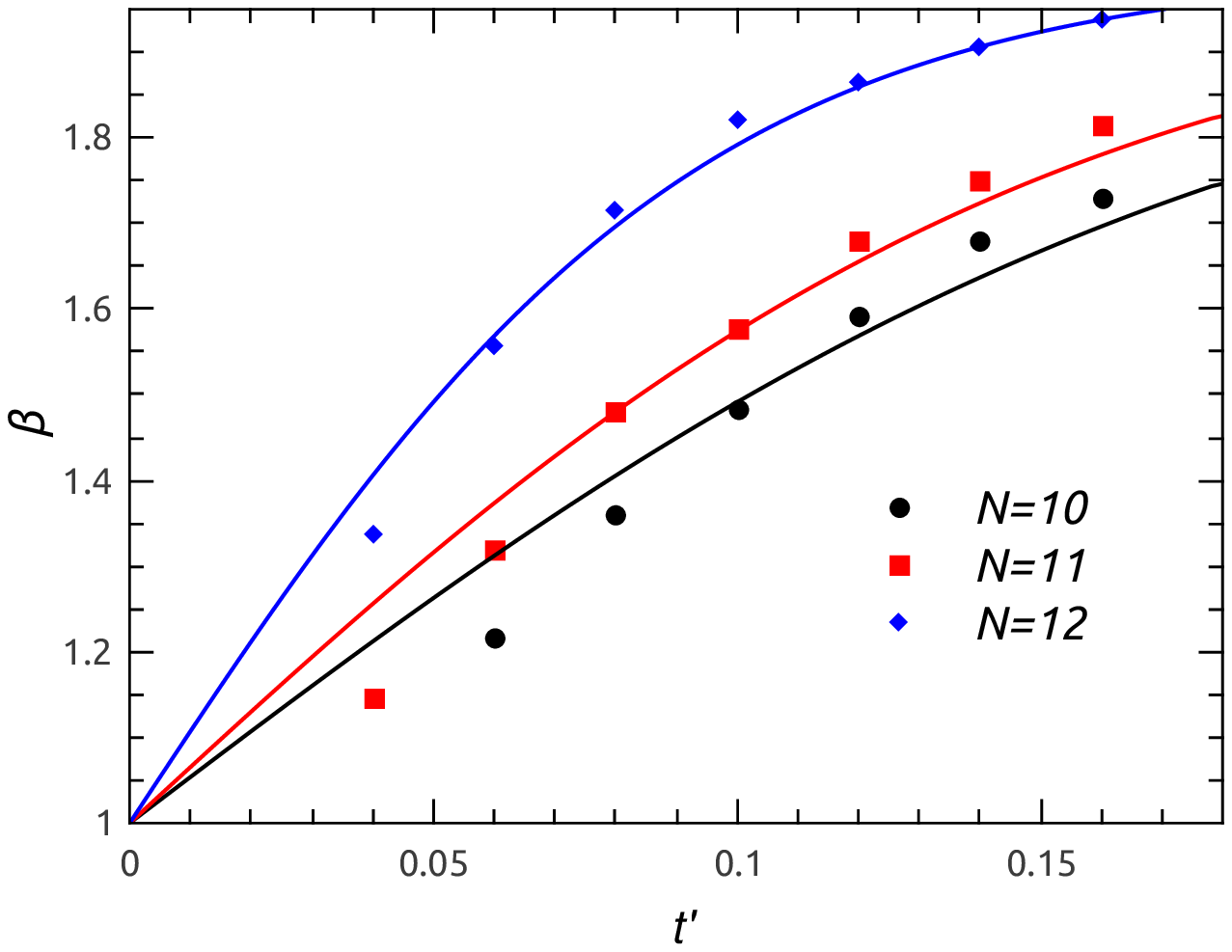}
\end{tabular}

\caption{
The change in level spacing distribution(A) for model I showing a crossover from GOE to GUE for $N=12$ (at half filling)
with $t=-1$ and $V=2$ and $h=4$ as
we $t'$ is increased. The dashed line is a fit to the GOE distribution and the solid line to GUE.
(B) The variation of the parameter $\beta$ with increasing $t'$ for $N=10,11,12$ for model I where $t=-1$,$V=2$ and $h=4$ and fit to the
function :$1+\tanh(t'/t'_{cr})$}
\label{Fig:level spacing GOE-GUE}
\end{figure}

\begin{eqnarray}
 P(s)=2c^{\beta+1}d^{-\beta-2}s^{\beta}\exp(-s^{2}c^{2}d^{-2})
 \label{Eq:brodyGUEGOE}
\end{eqnarray}
where, $c=\Gamma{[1+\beta/2]}$ and $d=\Gamma{[(\beta+1)/2]}$. $\beta$ changes from 1 (GOE) to 2(GUE)
as we increase $t'$. As for the previous cases, plotting $\beta$ as a function of $t'$ and fitting to $1+\tanh(t'/t'_{cr})$,
we calculate the crossover value of the time reversal symmetry breaking parameter (here $t'$) for different
system sizes as shown in Figure~\ref{Fig:level spacing GOE-GUE}.

\subsection{Poissonian to GSE}
For model II increasing on site disorder strength ($h$) and SOC parameter ($\mu$) simultaneously
(here we have always assume $h$=$\mu$) a crossover from Poissonian to GSE level statistics can be obtained as shown in Figure~\ref{Fig:poi_gse_3D}. Once again $P(s)$ is fit to a Brody form:

\begin{figure}
\centering
\begin{tabular}{cc}
\includegraphics[width=3.0in]{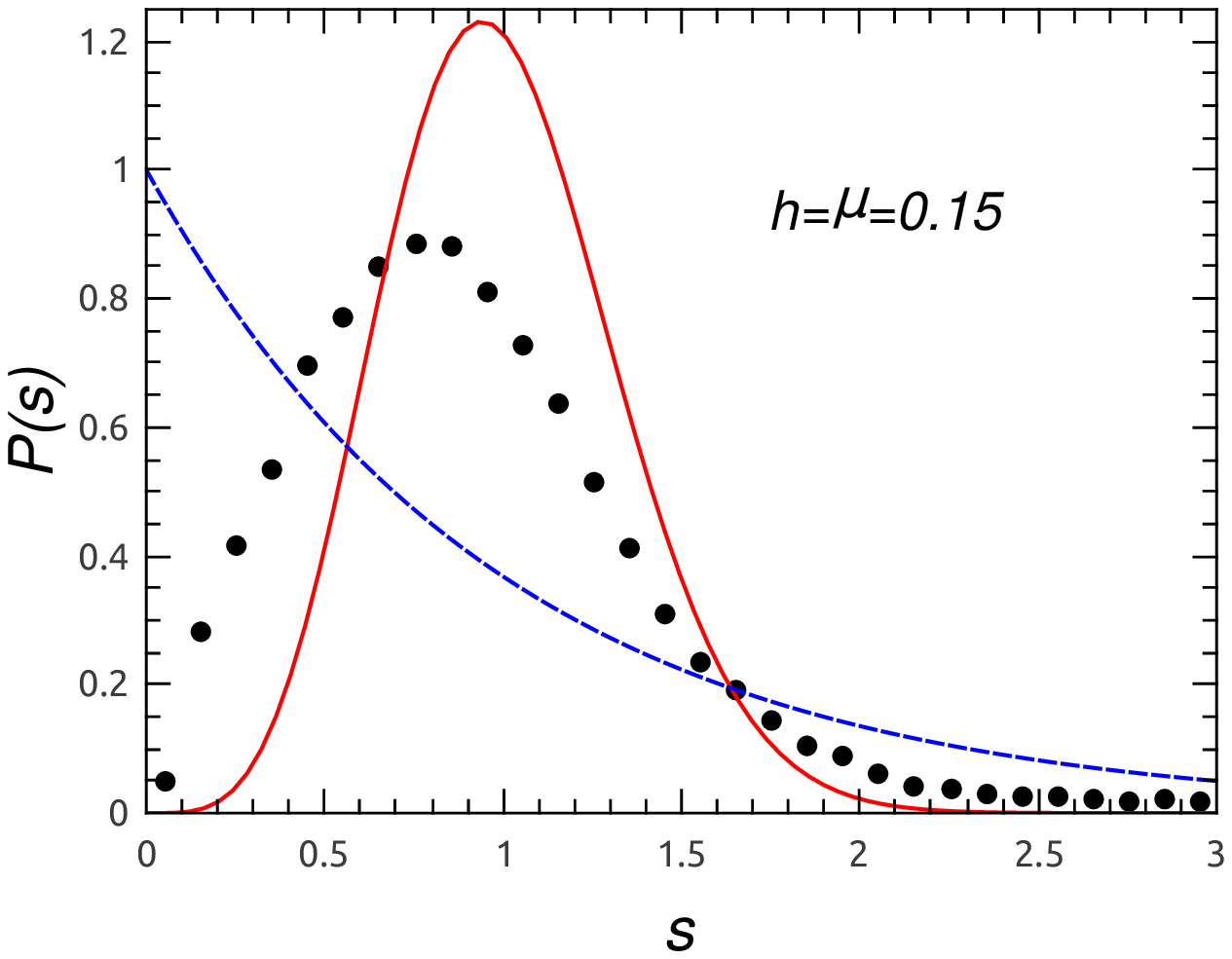} &
\includegraphics[width=3.0in]{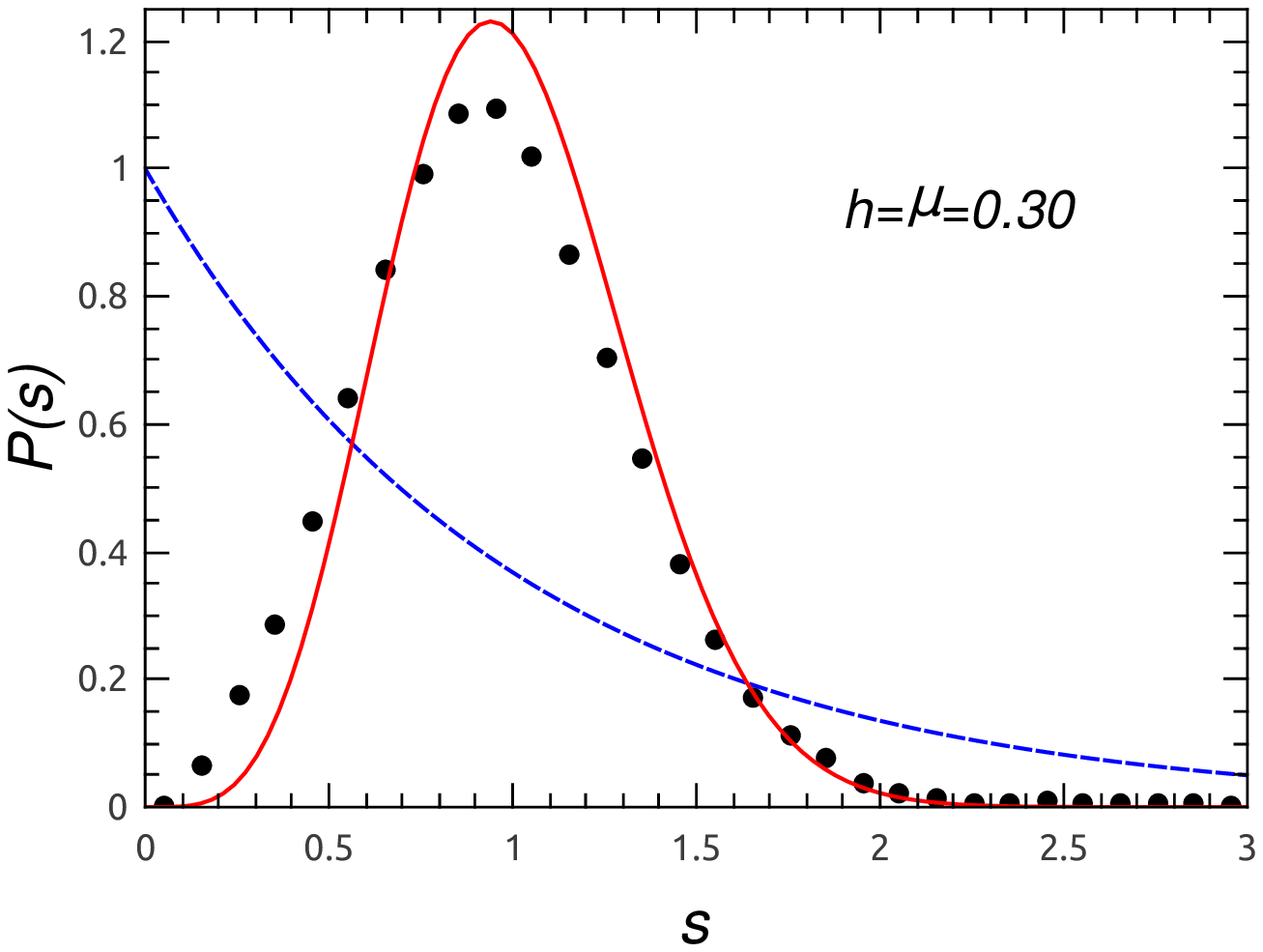} \\
\includegraphics[width=3.0in]{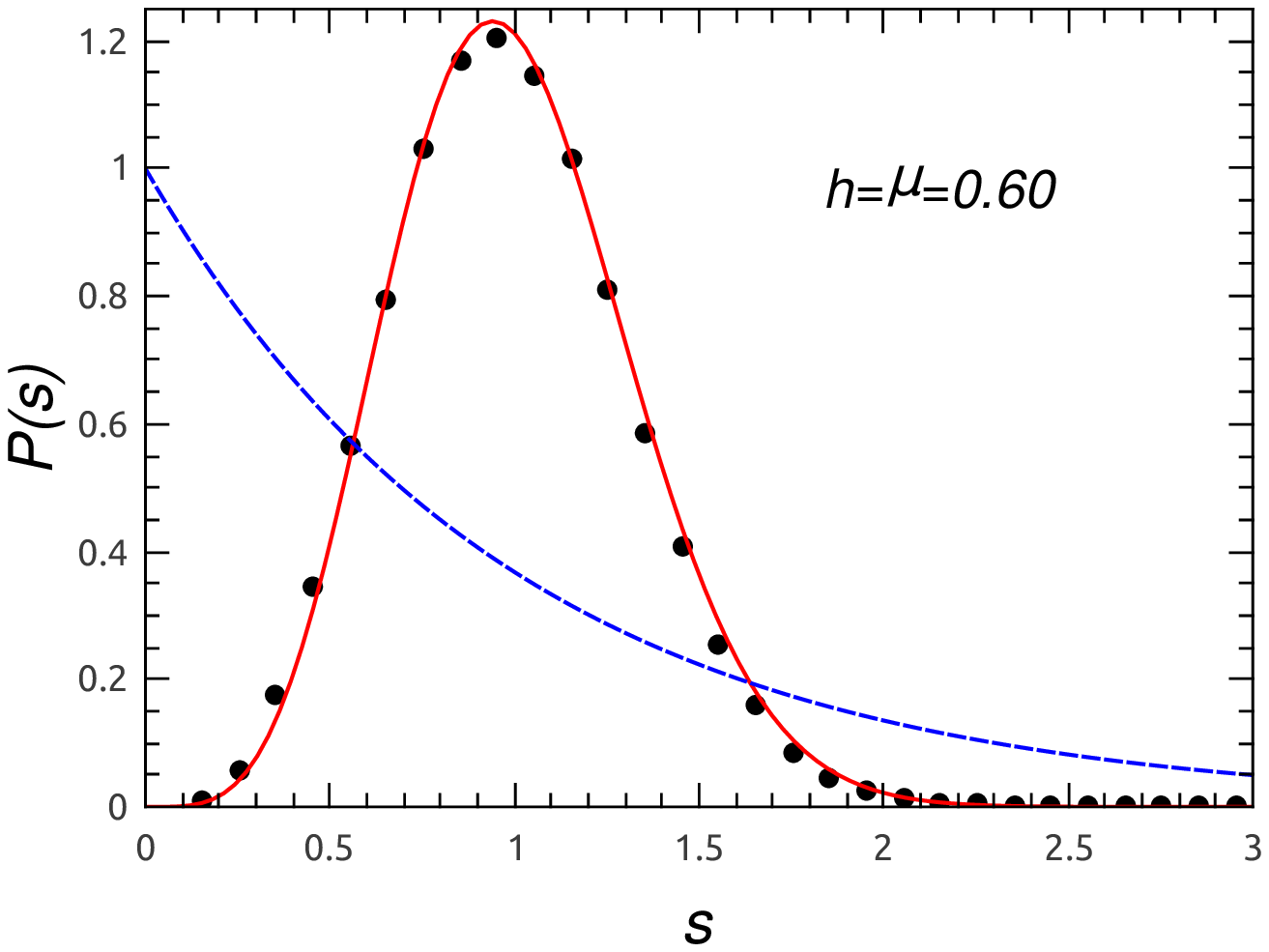} &
\includegraphics[width=3.0in]{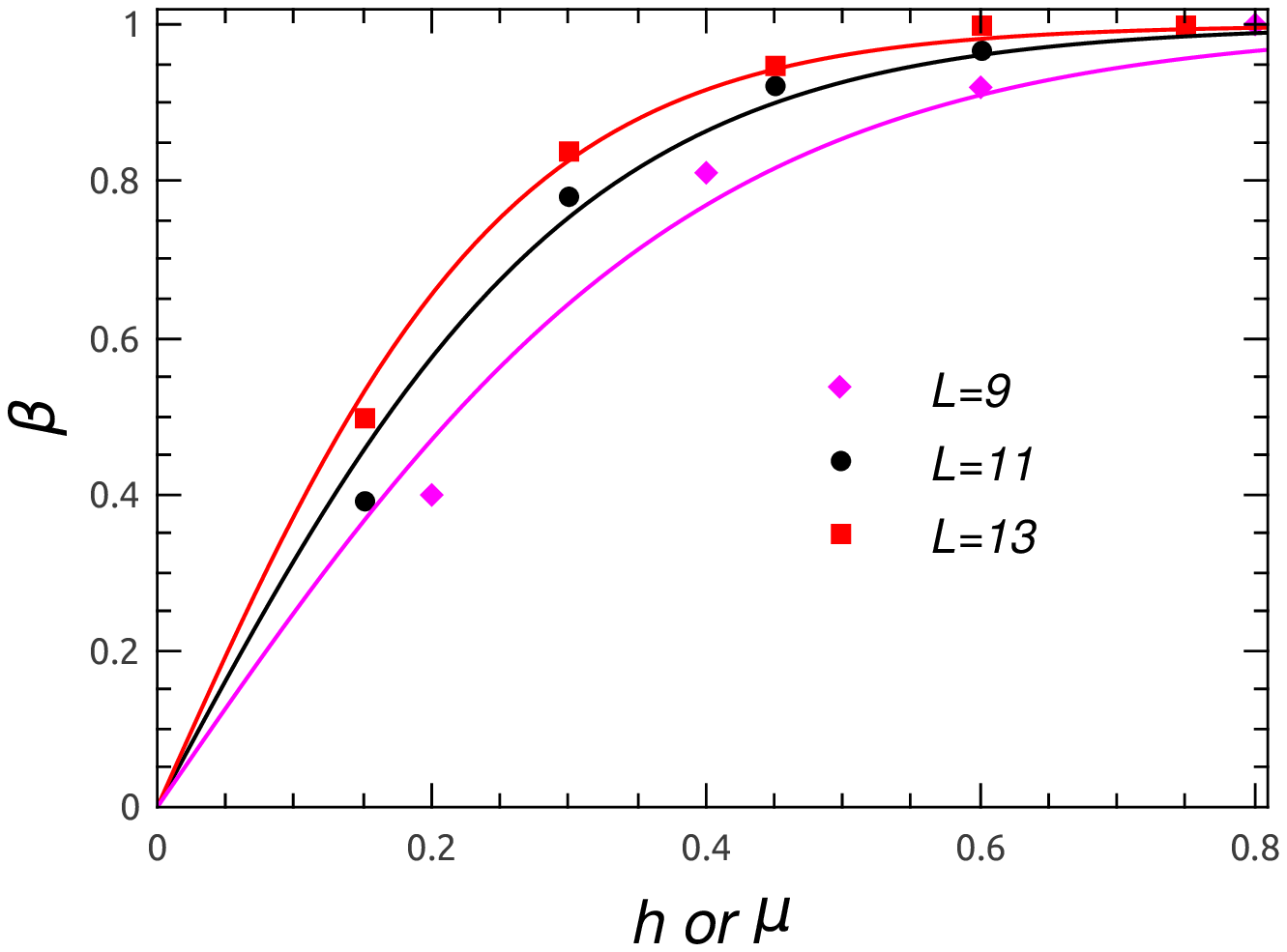}
\end{tabular}
\caption{(A)Level spacing distribution $P(s)$ for the 3D non-interacting disordered
model  for $L=11$. The values of the integrability
breaking parameters $h$ and $\mu$ are 0.15,0.3,0.6. The dashed line is a fit to the Poisson
distribution and the solid line to GSE.
(B)The variation of $\beta$ with $h$ (or $\mu$) and fit to the function $\tanh(h/h_{cr})$
 for $L$=9,11 and 13}
\label{Fig:poi_gse_3D}
\end{figure}

\begin{figure}
\centering
\begin{tabular}{cc}
\includegraphics[width=3.0in]{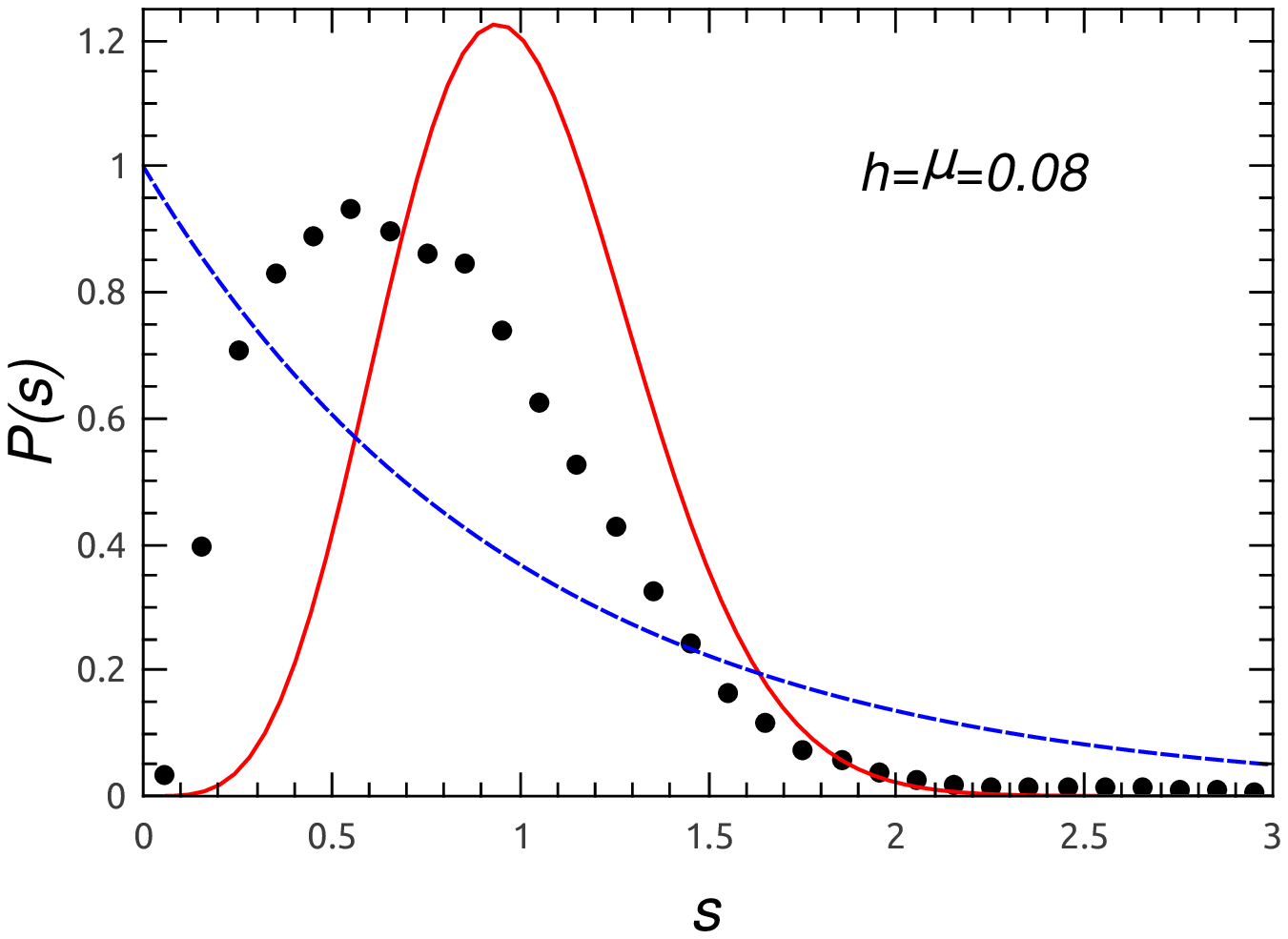} &
\includegraphics[width=3.0in]{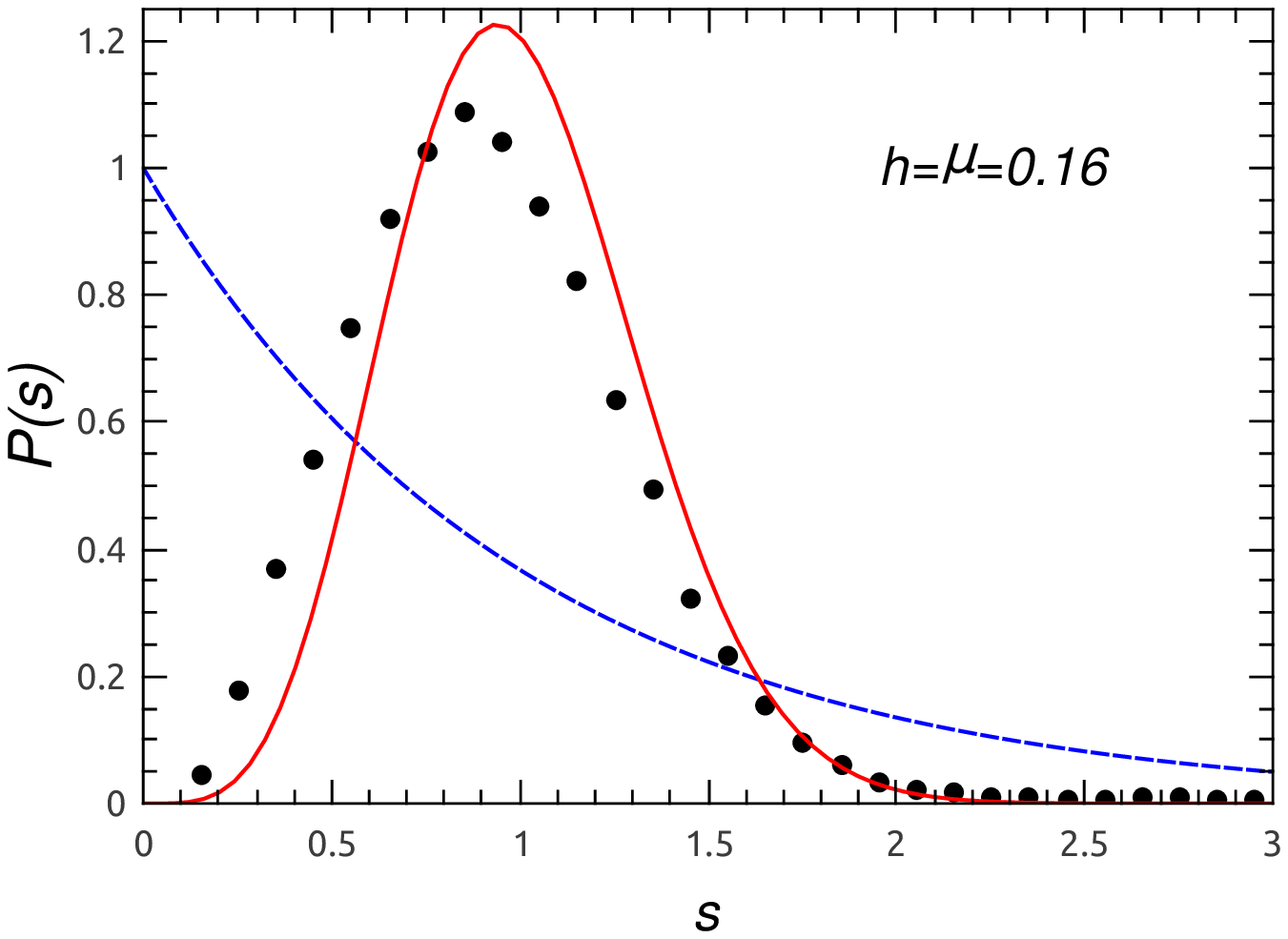} \\
\includegraphics[width=3.0in]{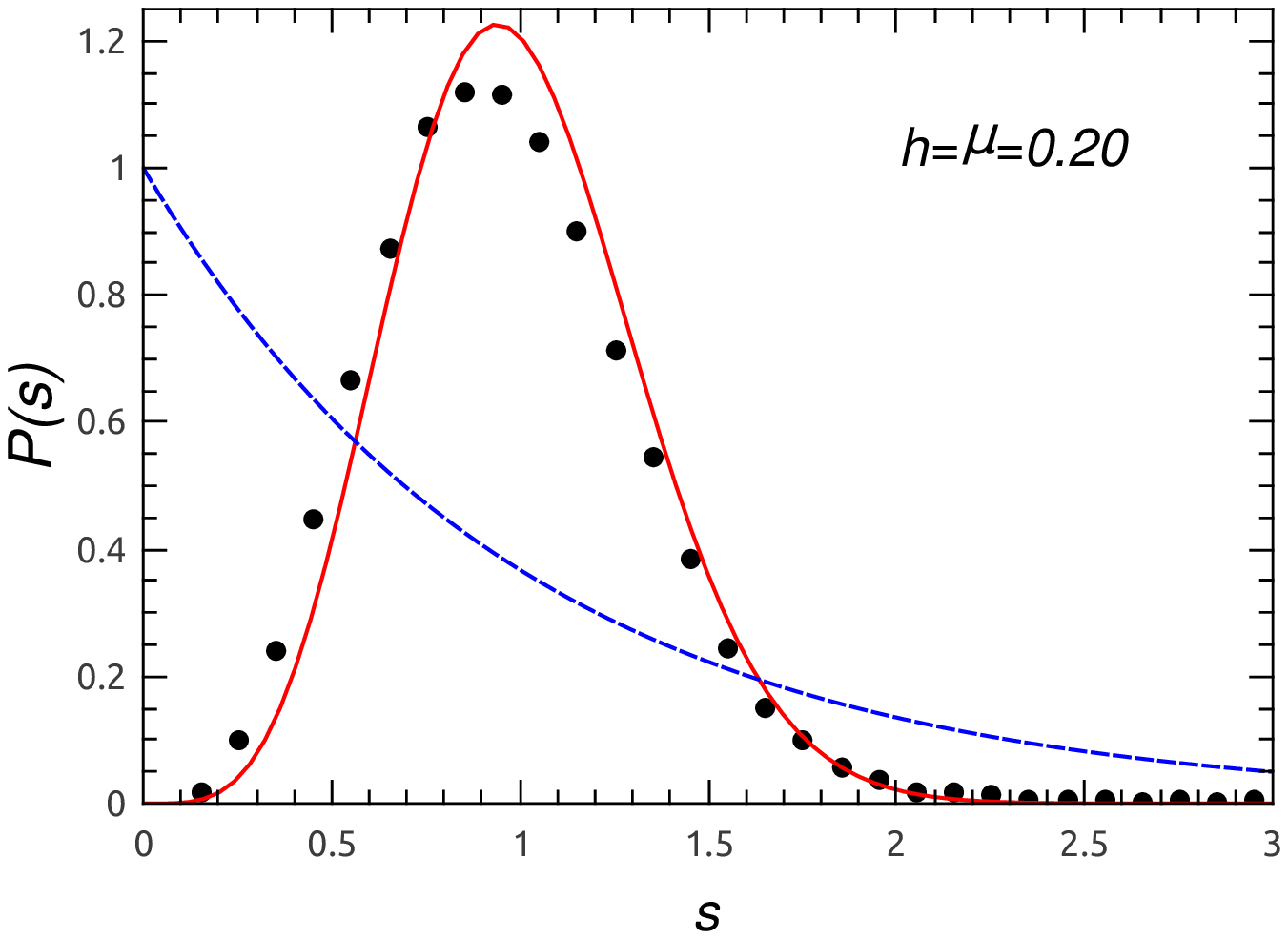} &
\includegraphics[width=3.0in]{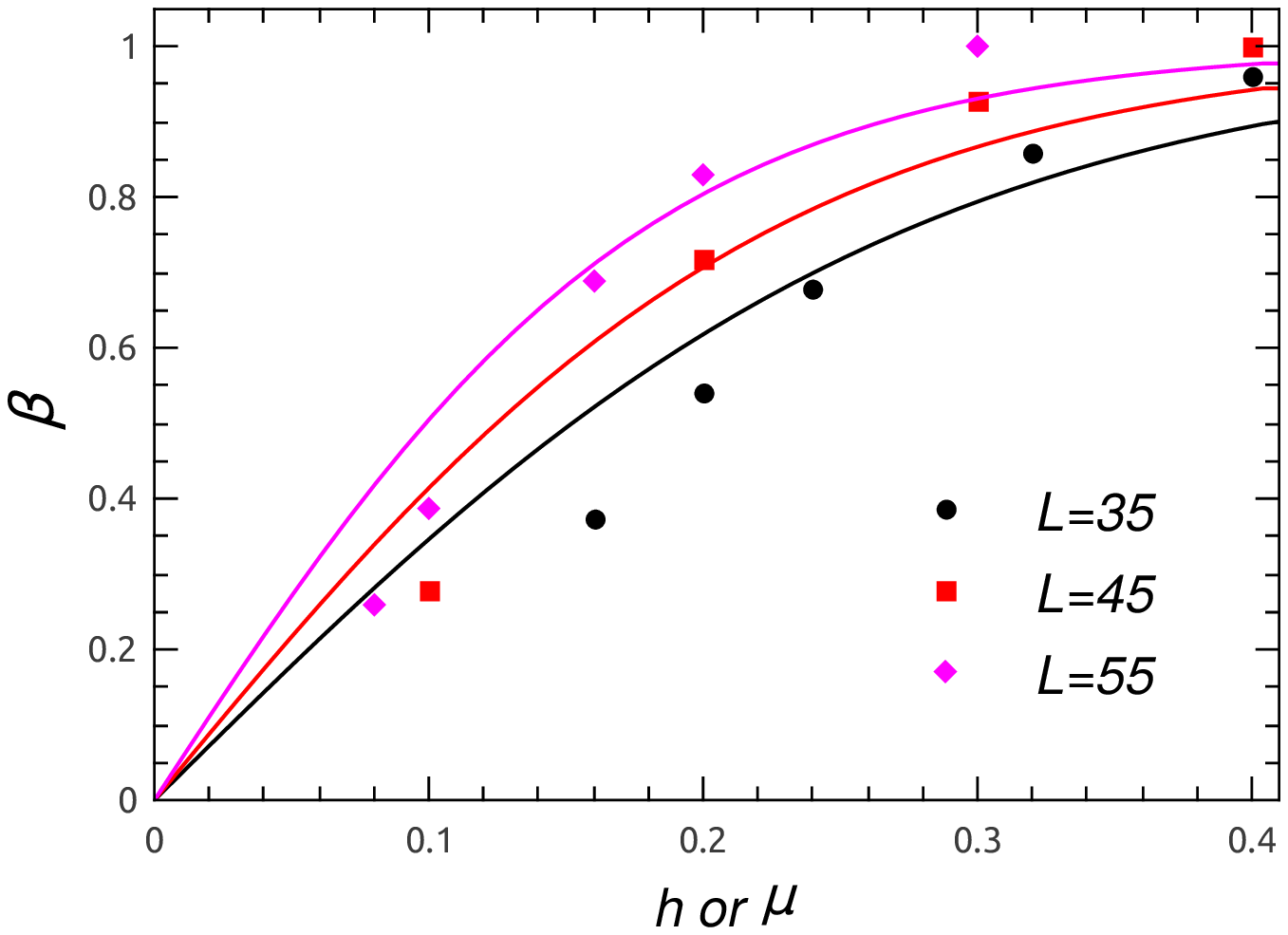}
\end{tabular}
\caption{(A)Level spacing distribution $P(s)$ for the 2D non-interacting disordered
model  for $L=55$. The values of the integrability
breaking parameters $h$ and $\mu$ are 0.08,0.16,0.20. The dashed line is a fit to the Poisson
distribution and the solid line to GSE.
(B)The variation of $\beta$ with $h$ (or $\mu$) and fit to the function $\tanh(h/h_{cr})$
 for $L$=35,45 and 55}
\label{Fig:poi_gse_2D}
\end{figure}

\begin{eqnarray}
P(s)=\frac{1+\beta}{\Gamma{[\frac{1+4\beta}{1+\beta}]}}a^{1+4\beta}s^{4\beta}\exp(-a^{\beta+1}x^{\beta+1})
\label{Eq:brodyGSE}
\end{eqnarray}
where, $a=\frac{\Gamma[\frac{2+4\beta}{1+\beta}]}{\Gamma[\frac{1+4\beta}{1+\beta}]}$. One can see a crossover from Poissonian to GSE level spacing statistics as
one changes $\beta$ from 0 to 1. Once again $\beta$ is plotted as a function
of $h$ (or $\mu$) and fit to the function $\tanh(h/h_{cr})$ to obtain the crossover value of disorder strength for different system sizes.

We have also investigated the Poissonian to GSE crossover for model II for a two dimensional square lattice. The change in level spacing distribution and variation of $\beta$ with $h$ or $(\mu)$ are shown in Figure~\ref{Fig:poi_gse_2D}.

\subsection{GOE to GSE}
\begin{figure}
\centering
\begin{tabular}{cc}
\includegraphics[width=3.0in]{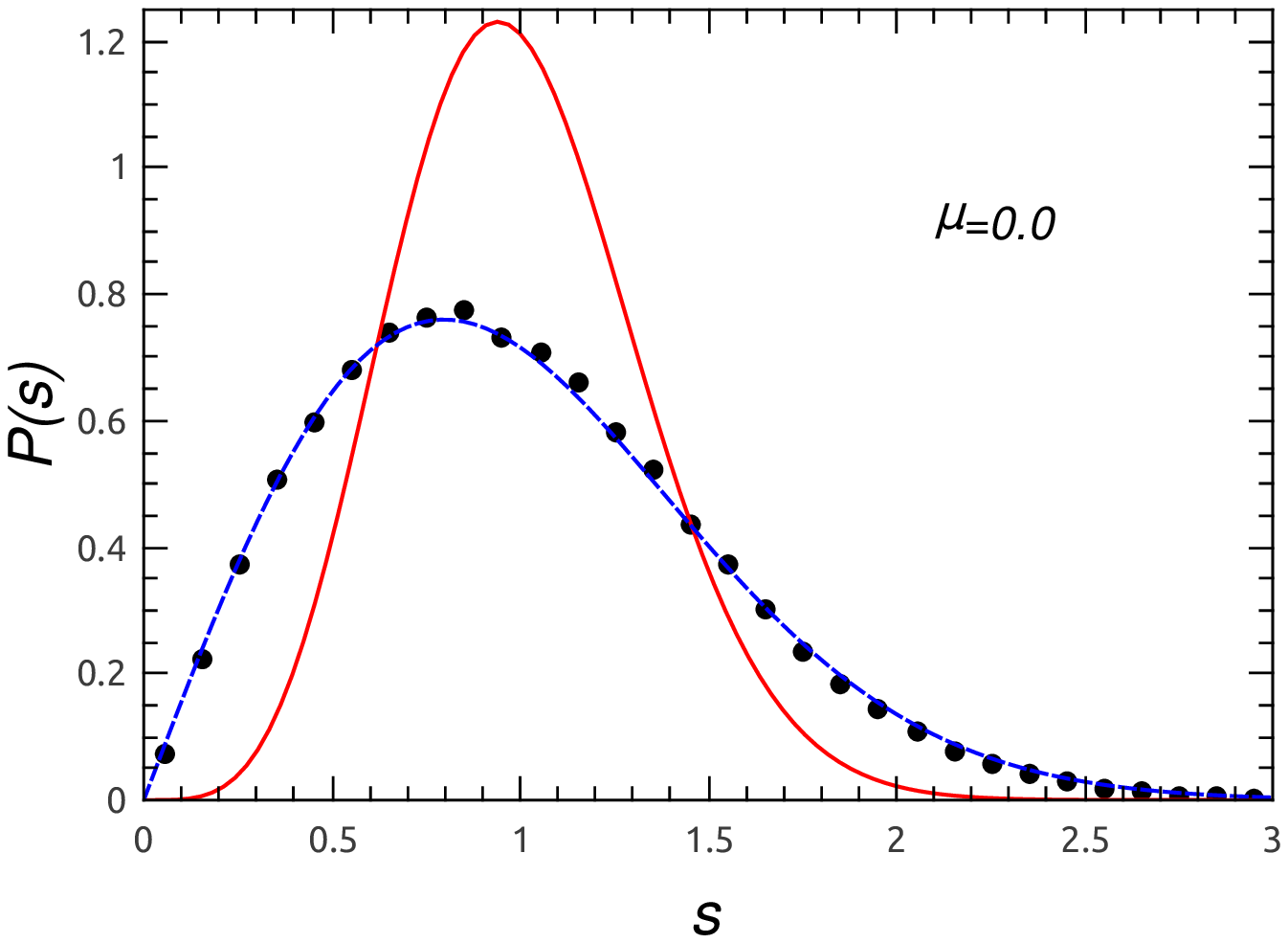} &
\includegraphics[width=3.0in]{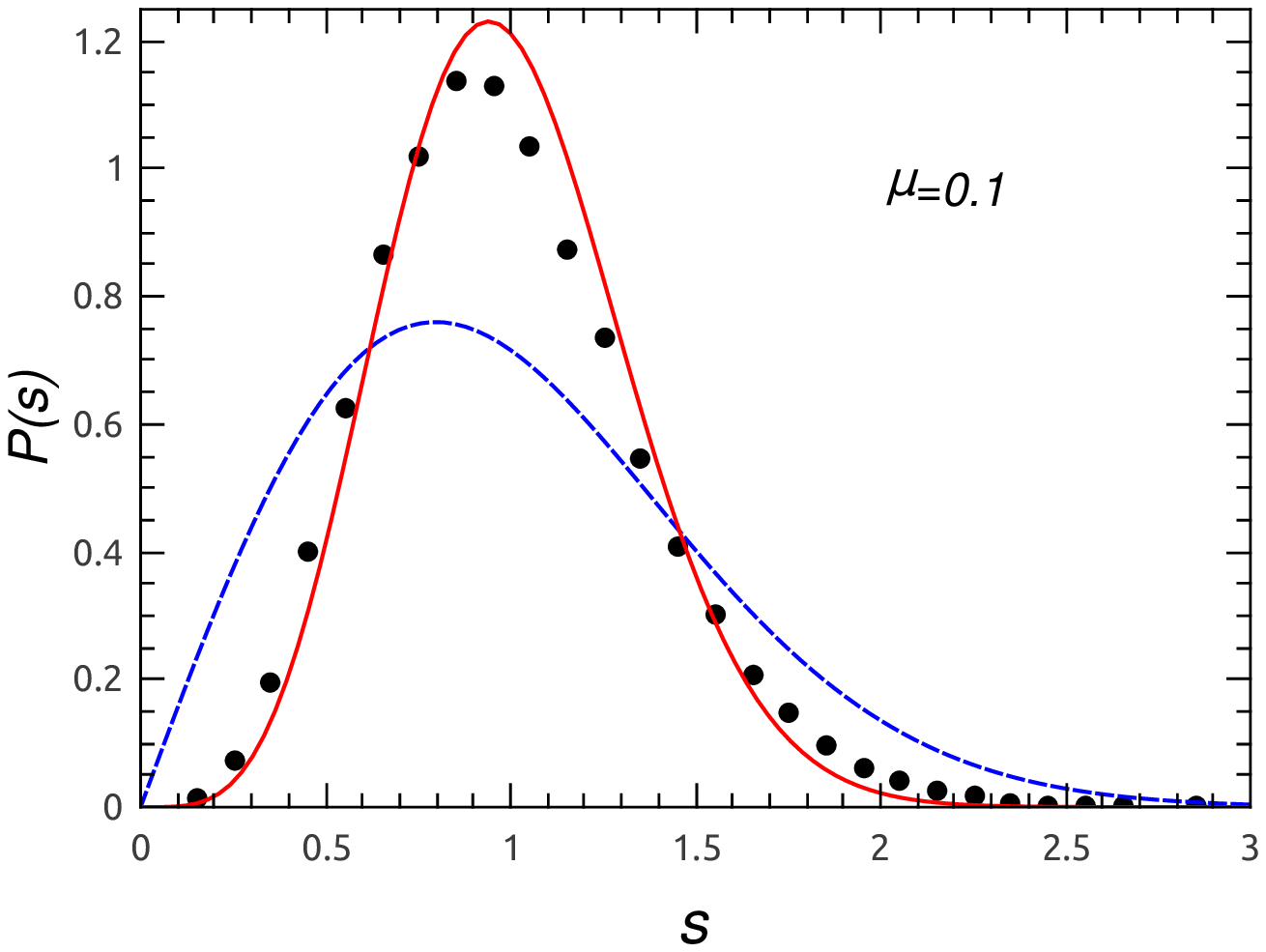} \\
\includegraphics[width=3.0in]{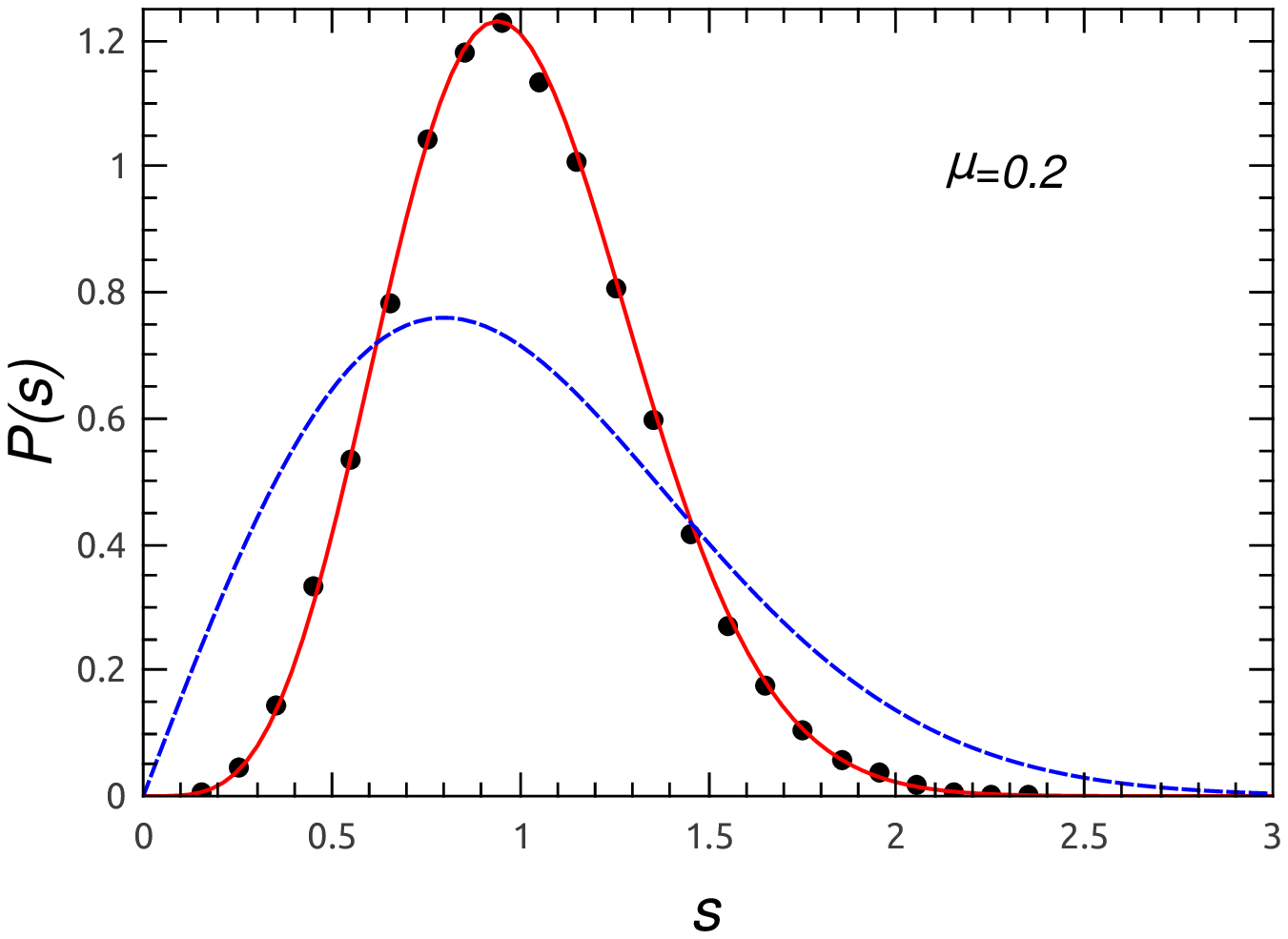} &
\includegraphics[width=3.0in]{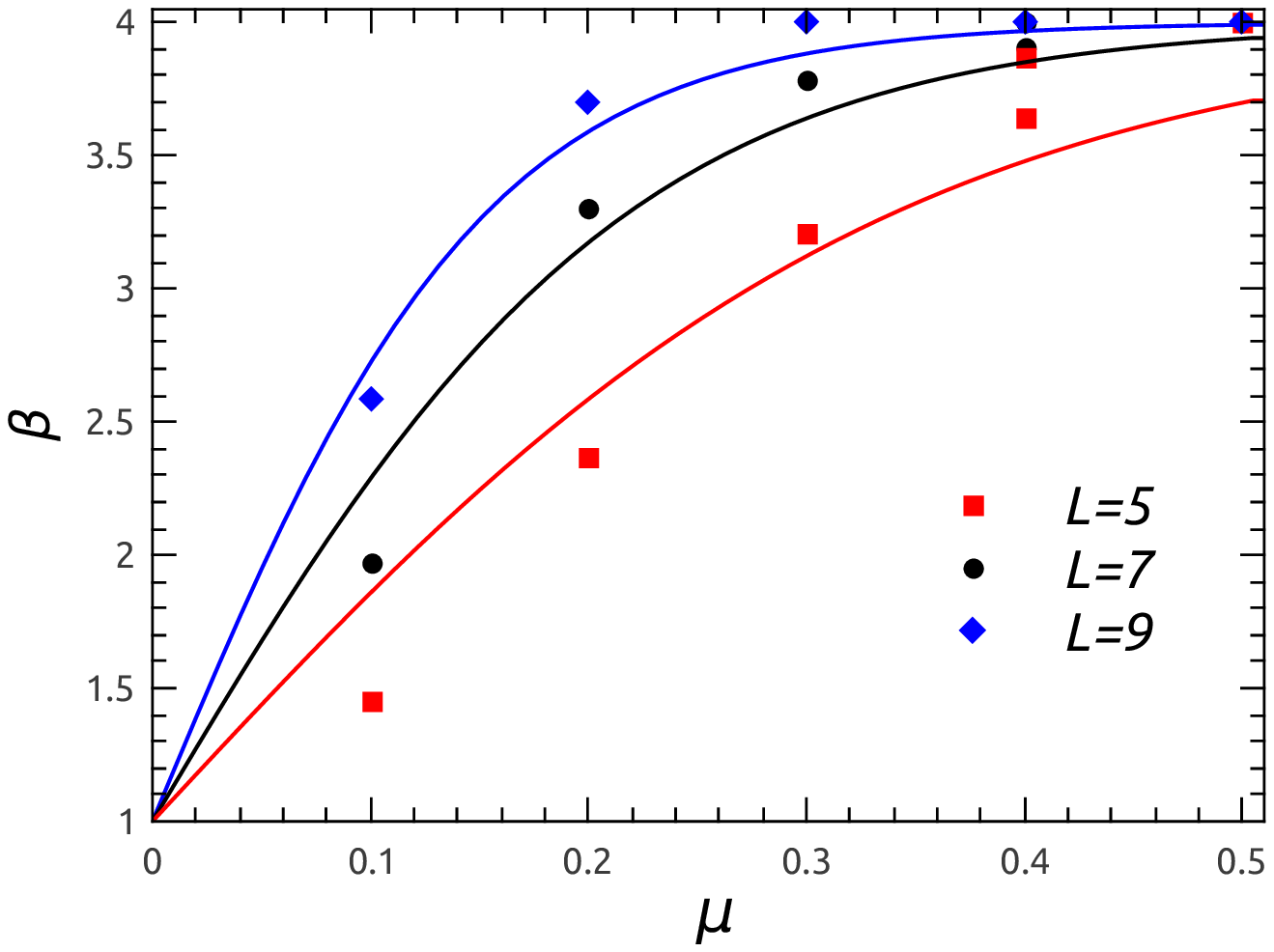}
\end{tabular}
\caption{(A)Level spacing distribution $P(s)$ for the 3D non-interacting disordered
model  for $L=11$. The values of the spin-rotation symmetry
breaking parameter  $\mu$ are 0.0,0.1,0.2. The dashed line is a fit to the GOE level spacing distribution
and the solid line to GSE.
(B)The variation of $\beta$ with $\mu$ and fit to the function :$1+3\tanh(\mu/\mu_{cr})$
 for $L$=5,7 and 9}
\label{Fig:goe_gse_3D}
\end{figure}

In the case of Model II if we set the SOC strength $\mu$ to zero and on site disorder $h$ to a finite value (here $h=4$), the level spacing
distribution will be GOE. Increasing $\mu$, we can break spin rotation symmetry and which will cause a GOE to GSE
crossover as observed in Figure~\ref{Fig:goe_gse_3D}. the Brody distribution in this case is the same as in Eqn.~\ref{Eq:brodyGUEGOE} except that $\beta$ can take any value between 1 (GOE) to 4 (GSE). The
variation of $\beta$ with $\mu$ is fit to the function $1+3\tanh(\mu/\mu_{cr})$ to calculate the crossover value
of $\mu$ for different system sizes.

\section{Results and discussion}
\begin{figure}
\centering
\begin{tabular}{cc}
\includegraphics[width=3.0in]{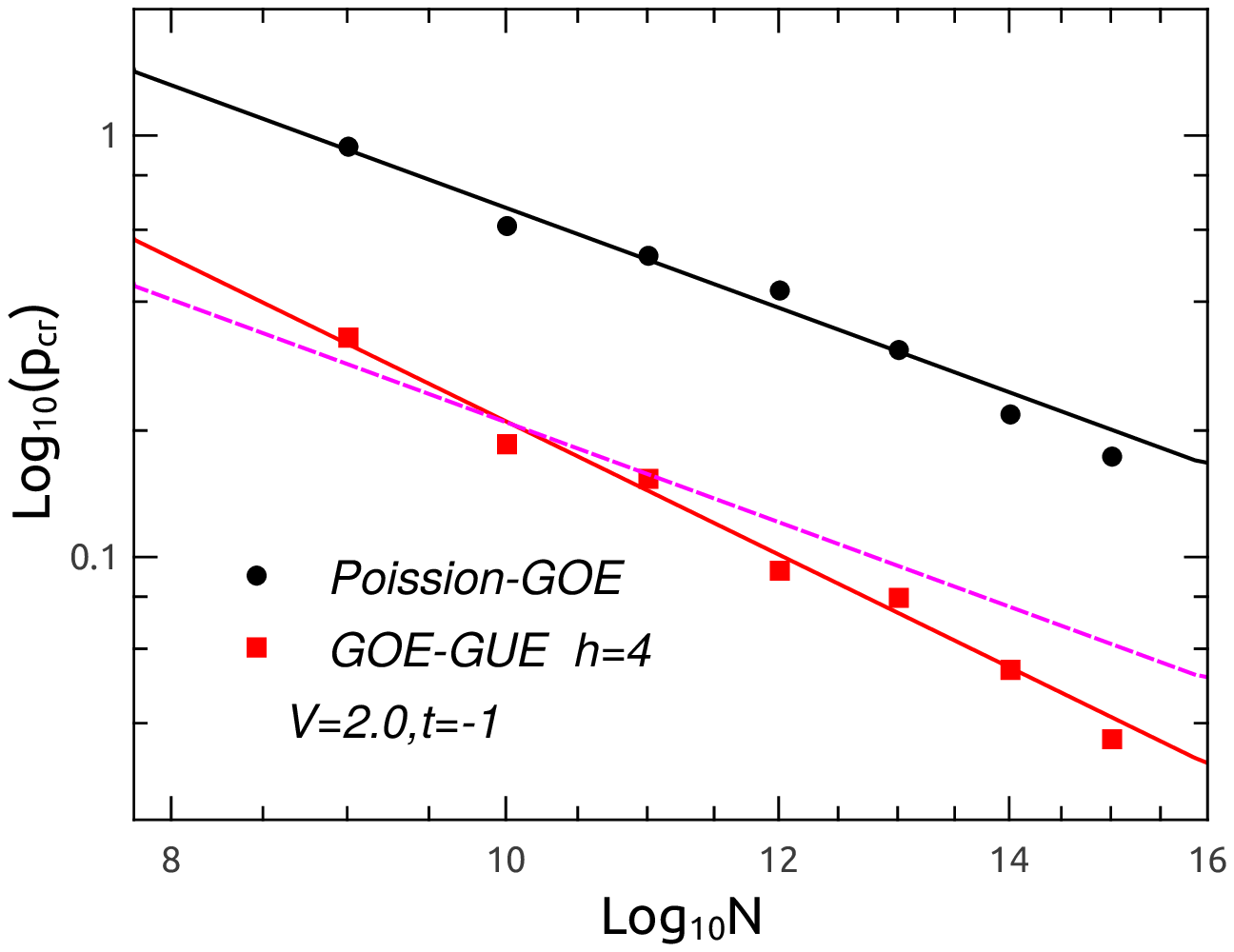} &
\includegraphics[width=3.0in]{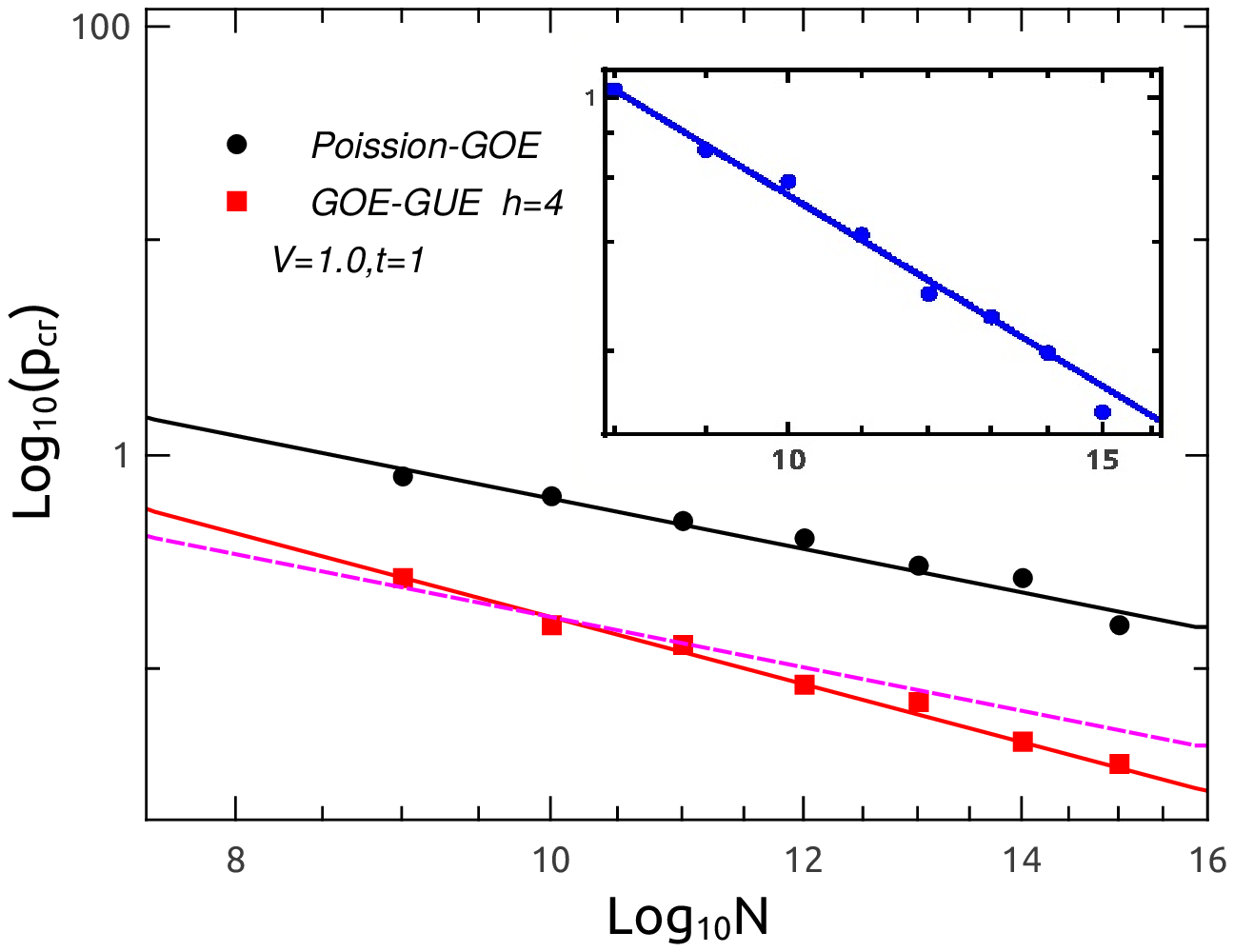} \\
\includegraphics[width=3.0in]{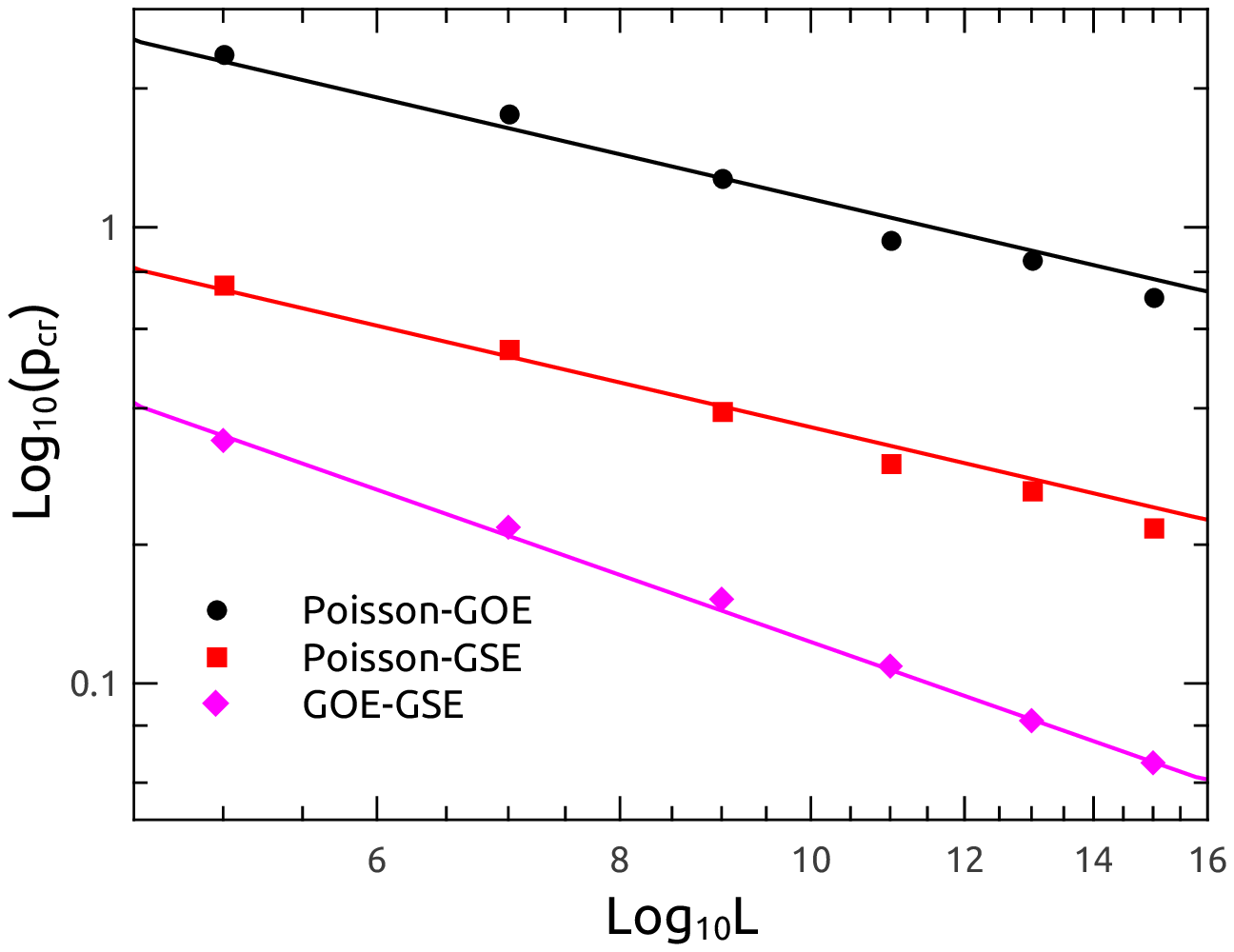} &
\includegraphics[width=3.0in]{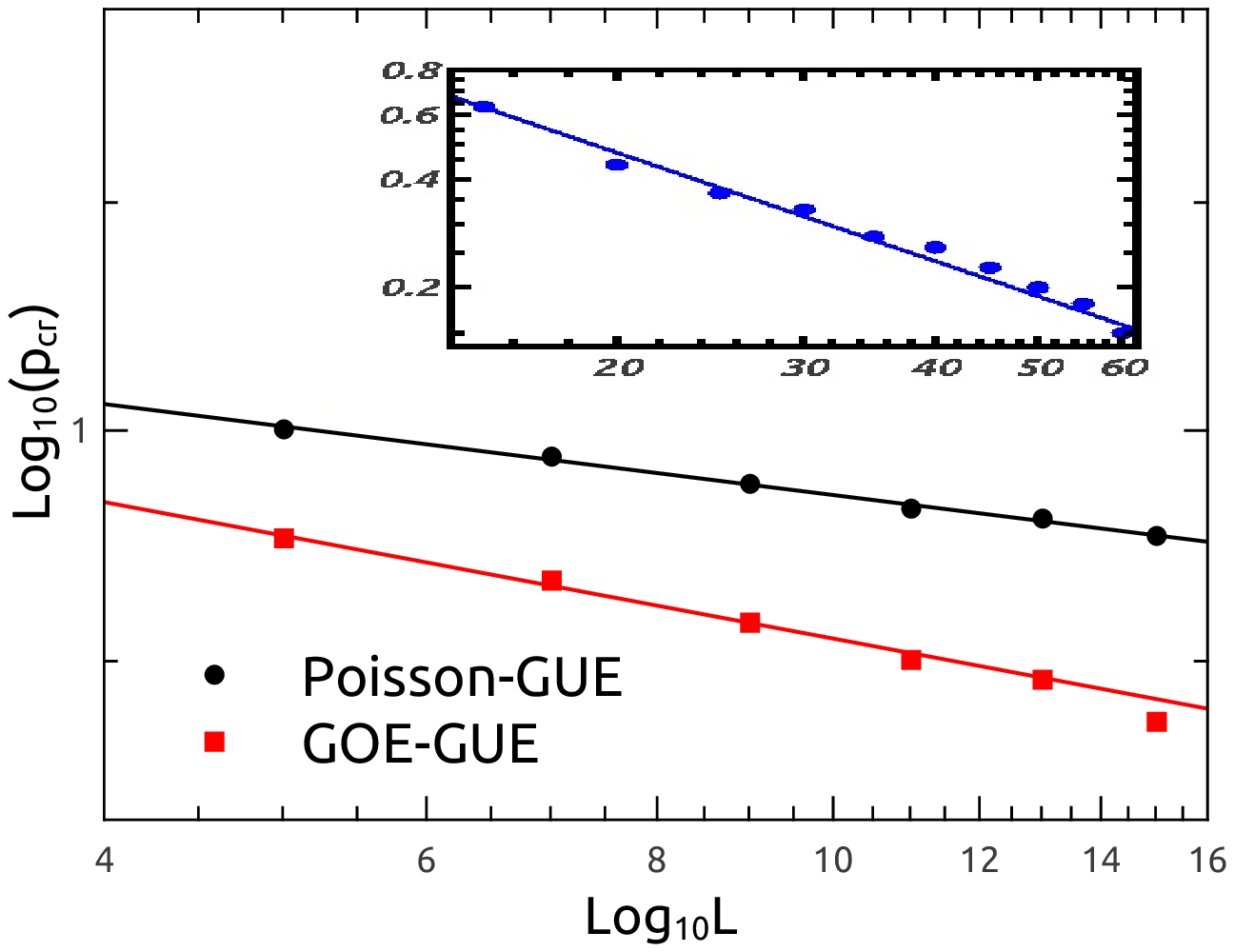}
\end{tabular}
\caption{Crossover values ($p_{cr}$) are plotted with system size on a log-log scale.
 For the one dimensional interacting model the exponent for the Poissonian to GOE and GOE to GUE crossovers are respectively
3 and 4. The dashed line is the best fit to the data with slope 3 for the GOE to GUE crossover. In the inset the value of the perturbation that causes the Poissonian-GUE crossover plotted against system size and the solid line corresponds to slope 3 when (A) $t=-1$,$V=2$ and
(B) $t=V=1$.
 For the non-interacting 3D disordered model (C) the perturbation causing the Poissonian-GOE and Poissonian-GSE scales as $~1/L$ and the one causing the
GOE to GSE crossover scale as $~1/L^{3/2}$, (D) the perturbation causing the Poissonian-GUE and GOE-GUE crossovers scale as $~1/L$ and
 $~1/L^{3/2}$ respectively. In the inset, the perturbation causingg the Poissonian-GSE crossover for the 2D model scales as $1/L$ with system size.
}

\label{Fig:level powerlaw goe-gue and poi-goe}
\end{figure}

We now summarize the results of our calculations. For Model I, we have found that the scale of the perturbation that causes a crossover from Poissonian to GOE and GUE level statistics goes to zero with increase of
system size as a power law with exponent 3 but for a GOE to GUE crossover the perturbation goes to zero
as a power law with exponent 4.(Figure~\ref{Fig:level powerlaw goe-gue and poi-goe})
We also obtain the exponent 3 for the Poisson to GOE crossover also for the interacting 1D system without disorder~\cite{modak.2013}.

For Model II, we have observed that the integrability breaking parameter which is responsible for the Poissonian to
GOE/GSE crossover goes to zero as a power law with exponent 1 and for the GOE to GSE
crossover this exponent becomes 3/2.(Figure~\ref{Fig:level powerlaw goe-gue and poi-goe})

We have also studied a different version of model II to investigate the Poisson to GUE and GOE to GUE crossovers.
Here, we have taken the nearest neighbor hoping matrix $V_{ij}$ to have the following form,
\begin{eqnarray}
V_{i,j}=
 \quad
\begin{pmatrix}
1+i\mu V^{z} & 0 \\
0 & 1+i\mu V^{z}
\end{pmatrix}
\quad
\end{eqnarray}
This model breaks time reversal symmetry. Increasing $\mu$ and $h$ simultaneously we have observed a Poissonian-GUE
crossover and then setting $h=4$ and increasing $\mu$, a GOE to GUE crossover. The exponents of the power law scaling of the crossover value of the perturbation are 1 and 3/2 for the Poissonian-GUE and GOE-GUE cases respectively.

Why do we obtain the same exponent for the Poissonian to GUE and Poissonian to GSE crossovers as we do
for Poissonian to GOE? A possible answer is that the Poissonian to GUE and Poissonian to GSE crossovers can be thought of as a crossover first from Poissonian to GOE, which introduces level repulsion and then a crossover from GOE to GUE and GSE through the breaking of additional symmetries (time reversal for GOE-GUE and spin rotation invariance for GOE-GSE). For this interpretation to be correct, the fall off of the crossover value with system size for GOE-GUE and GOE-GSE has to be faster than for Poissonian-GOE. This is indeed the case for all the cases we have considered as can be seen from Figure~\ref{Fig:level powerlaw goe-gue and poi-goe}.

A second important question is why the Poissonian-GOE/GUE/GSE crossover scaling exponents are different for the two different models? The reason does not seem to be that the models have different dimensionality. For instance, the Poissonian-GOE crossover for model II is the same in both two and three dimensions. We conjecture that this difference can be attributed to the fact that model I contains interactions among the elementary degrees of freedom while model II does not. This has a bearing on the integrable limits of these models as well in that the elementary degrees of freedom are correlated even in the integrable limit of model I and not model II. This difference could well be responsible for the different exponents in the two models. To further investigate this and also the role of dimensionality, it would be desirable to obtain crossovers among all the different categories of ensembles in different dimensions. However, it is not possible for us to realize a GOE phase in the non-interacting model
in two dimensions owing to the fact that localization becomes operative in the absence of spin-orbit coupling and time reversal breaking in two dimensions. On the other hand, it is difficult to study model I in more than one dimension since the presence of interactions greatly reduces the system sizes we can study in higher dimensions. A through investigation of the effect of interactions and dimensionality on the crossovers among different ensembles will require a study of more models which we will defer to a later work.

\section{Conclusions}
We have investigated the finite scaling of perturbations which cause crossovers among different random matrix ensembles in two different models: a one dimensional model of interacting hard core bosons (or equivalently spin 1/2) objects and a disordered model of non-interacting particles with disorder and spin-orbit coupling. Obtaining the level spacing distribution using numerical exact diagonalization, we have found that the scaling is a power law for crossovers among all the different ensembles (Poissonian, GOE, GUE and GSE) that can be realized in these models and obtained the corresponding exponents. We find that the scaling from Poissonian to any of the other ensembles is dominated by the Poissonian-GOE crossover which introduces level repulsion while the symmetry breaking that causes the GOE-GUE and GOE-GSE crossovers is a sub-dominant effect. We also conjecture that there is a difference in the scaling depnding on whether the elementary degrees of freedom interact or not.

\section{Acknowledgments}
The authors would like to acknowledge many useful discussions with Sriram Ramaswamy. SM thanks the Department of Science and Technology, Government of India.
RM thanks Tapan Mishra for discussions and acknowledges support from UGC Fellowship.

\end{document}